\documentclass[aps,prd,reprint,superscriptaddress,nofootinbib]{revtex4-2}

\usepackage{mathrsfs,amsbsy,amssymb,latexsym,amsfonts,amsmath,amsthm}
\usepackage{graphicx}
\usepackage{dcolumn}
\usepackage{bm}

\newcommand{\bbC}{{\mathbb{C}}}
\newcommand{\bbR}{{\mathbb{R}}}

\newcommand{\bx}{{\textbf{x}}}
\newcommand{\by}{{\textbf{y}}}
\newcommand{\bz}{{\textbf{z}}}

\begin{document}


\preprint{KUNS-2764}

\title{Applying the tempered Lefschetz thimble method 
to the Hubbard model away from half-filling}


\author{Masafumi Fukuma}
\email{fukuma@gauge.scphys.kyoto-u.ac.jp}
\affiliation{%
 Department of Physics, Kyoto University,
 Kyoto 606-8502, Japan
}%
\author{Nobuyuki Matsumoto}%
\email{nobu.m@gauge.scphys.kyoto-u.ac.jp}
\affiliation{%
 Department of Physics, Kyoto University,
 Kyoto 606-8502, Japan
}%
\author{Naoya Umeda}
\email[]{naoya.umeda1134@gmail.com}
\affiliation{PricewaterhouseCoopers Aarata LLC, 
Otemachi Park Building, 1-1-1 Otemachi, Chiyoda-ku, Tokyo 100-0004, Japan}



\begin{abstract}
The tempered Lefschetz thimble method 
is a parallel-tempering algorithm towards solving the numerical sign problem. 
It uses the flow time of the gradient flow 
as a tempering parameter 
and is expected to tame both the sign and multimodal problems simultaneously. 
In this paper, we further develop the algorithm  
so that the expectation values can be estimated precisely 
with a criterion ensuring global equilibrium 
and the sufficiency of the sample size. 
To demonstrate that this algorithm works well, 
we apply it to the quantum Monte Carlo simulation 
of the Hubbard model away from half-filling 
on a two-dimensional lattice of small size, 
and show that the numerical results agree nicely with exact values.
\end{abstract}


\maketitle


\section{Introduction
\label{sec:introduction}}
The sign problem is one of the major obstacles 
when performing numerical calculations in various fields of physics. 
Typical examples include 
finite density QCD \cite{Aarts:2015tyj}, 
quantum Monte Carlo (QMC) calculations of quantum statistical systems 
\cite{Hirsch:1985,Loh:1990,Pollet:2012}, 
and the numerical simulations of real-time quantum field theories.

Among a variety of approaches, 
two algorithms have taken attention 
as potential candidates to generically solve the sign problem 
for systems with complex action; 
one is the complex Langevin method \cite{Parisi:1984cs}, 
and the other is a class of algorithms utilizing the Lefschetz thimbles 
\cite{Cristoforetti:2012su,Cristoforetti:2013wha,Mukherjee:2013aga,
Fujii:2013sra,Cristoforetti:2014gsa,Alexandru:2015xva,
Alexandru:2015sua,Alexandru:2016ejd,
Fukuma:2017fjq,Alexandru:2017oyw}. 
Although both the algorithms make use of complexification of variables 
and analytic continuation of integrands, 
their methodologies are fairly different; 
the former algorithm attempts to replace the complex Boltzmann weight 
by a real positive weight defined in the whole complex space, 
while the latter deforms the integration region in the complex space 
so as to reduce the phase oscillation. 
At this stage, each algorithm has its own advantage and disadvantage. 
The former is advantageous in that it is relatively fast 
with computational cost $O(N)$ ($N$: the degrees of freedom), 
but it suffers from the so-called wrong convergence problem 
\cite{Ambjorn:1985iw,Aarts:2011ax,Aarts:2013uxa,Nagata:2016vkn}. 
The latter is generally free from the wrong convergence problem 
{\em if} only a single thimble is relevant 
in evaluating the expectation values of physical observables of interest. 
The disadvantage is its expensive numerical cost, 
which is $O(N^3)$ because of the need to calculate 
the Jacobian determinant. 
When multiple thimbles are relevant, 
one needs to take care of the multimodality of the distribution. 
The tempered Lefschetz thimble method (TLTM) 
was thus proposed in \cite{Fukuma:2017fjq} 
to tame both the sign and multimodal problems simultaneously, 
where the system is tempered 
by the flow time of the antiholomorphic gradient flow 
(see also \cite{Alexandru:2017oyw} for a similar idea).

In this paper, we further develop the TLTM, 
proposing an algorithm 
which allows the precise estimation of expectation values 
with a criterion ensuring global equilibrium 
and the sufficiency of the sample size. 
The key is the use of the fact 
that the expectation values should be the same for all flow times. 
To demonstrate that this algorithm works well, 
we apply it to the QMC simulation 
of the Hubbard model away from half-filling. 

The application of Lefschetz thimble methods to the Hubbard model 
has already been considered by several groups 
\cite{Mukherjee:2014hsa,Tanizaki:2015rda,Ulybyshev:2017hbs} 
(see also \cite{Ulybyshev:2019a,Ulybyshev:2019b} for recent study), 
and the relevance of the contributions from multiple thimbles 
has been reported. 
In this paper, 
we consider a two-dimensional periodic square lattice 
of size $N_s=2\times 2$ 
with the inverse temperature decomposed to $N_\tau=5$ pieces, 
and numerically evaluate the expectation values of observables 
as functions of the chemical potential 
with other parameters fixed to some values. 
We show that the TLTM 
(the implementation of tempering 
combined with the above algorithm for precise estimation) 
give results that agree nicely with exact values, 
simultaneously resolving the sign and multimodal problems.

We comment that 
the extent of seriousness of the sign problem 
in the QMC simulation of the Hubbard model 
depends heavily on the choice of the Hubbard-Stratonovich variables. 
In this paper, 
in order to apply the Lefschetz thimble method,
we exclusively consider a Gaussian Hubbard-Stratonovich variable 
that leads to a complex action. 
There the sign problem is actually severe as we will see below, 
and one needs to seriously consider 
a dilemma between the sign and multimodal problems, 
which can be solved by the TLTM as stated above. 
However, the temporal size considered here is still small ($N_\tau=5$), 
and for such a high temperature regime 
one can resort to other methods than the Lefschetz thimble methods 
with a different type of Hubbard-Stratonovich variables 
(see discussions in section \ref{sec:Hubbard}).

This paper is organized as follows. 
In section \ref{sec:TLTM} 
after briefly reviewing the TLTM \cite{Fukuma:2017fjq}, 
we give a new algorithm 
which allows the precise estimation of expectation values 
with a criterion ensuring global equilibrium 
and the sufficiency of the sample size. 
This algorithm is applied to the Hubbard model 
in section \ref{sec:Hubbard}, 
and we discuss about the obtained numerical results. 
We there also make a comment on the sign averages 
obtained by other methods. 
Section \ref{sec:conclusion} is devoted 
to conclusion and outlook. 
Five appendices are given for more detailed discussions on various topics.

\section{Tempered Lefschetz thimble method
\label{sec:TLTM}}
Let $x=(x^i)\in\bbR^N$ be a real $N$-dimensional dynamical variable 
with action $S(x)$ which may take complex values. 
Our main concern is to estimate the expectation values 
\begin{align}
 \langle \mathcal{O}(x) \rangle_S 
 \equiv \frac{\int_{\bbR^N} dx\, e^{-S(x)}\,\mathcal{O}(x)}
 {\int_{\bbR^N} dx\, e^{-S(x)}}.
\label{eq:vev1}
\end{align}
We assume that $e^{-S(z)}$ and $e^{-S(z)}\,\mathcal{O}(z)$ 
are entire functions over $\bbC^N$ 
when $x$ is complexified to $z=(z^i)\in\bbC^N$. 
Then, due to Cauchy's theorem for higher dimensions, 
the right-hand side does not change 
under continuous deformations of the integration region 
as long as the boundary at infinity is kept fixed 
so that the integrals converge. 
The sign problem will get reduced 
if ${\rm Im}\, S(z)$ is almost constant 
on the new integration region.

In \cite{Alexandru:2015xva,Alexandru:2015xva,
Alexandru:2015sua,Alexandru:2016ejd,
Fukuma:2017fjq,Alexandru:2017oyw} 
such a deformation $x \to z_t(x)$ ($t\geq 0$) is made 
according to the antiholomorphic gradient flow: 
\begin{align}
 \dot{z}_t^i &= [\partial_i S(z_t)]^\ast,
 \quad
 z^i_{t=0} = x^i. 
\label{eq:flow1}
\end{align}
Equation \eqref{eq:vev1} can then be rewritten as
\begin{align}
 \langle \mathcal{O}(x) \rangle_S 
 = \frac{\int_{\Sigma_t} d  z\, e^{-S(z)}\,\mathcal{O}(z)}
 {\int_{\Sigma_t} d z\, e^{-S(z)}}
 \quad (\Sigma_t \equiv z_t(\bbR^N)), 
\label{eq:LT1}
\end{align}
which can be further rewritten 
as a ratio of reweighted integrals over $\bbR^N$ 
by using the Jacobian matrix 
$J_t(x)\equiv \bigl(\partial z_t^i(x)/\partial x^j\bigr)$ 
\cite{Alexandru:2015xva}: 
\begin{align}
 \langle \mathcal{O}(x) \rangle_S 
 &= \frac{\int_{\bbR^N} d x\, 
 \det\!J_t(x)\,e^{-S(z_t(x))}\,\mathcal{O} (z_t(x))}
 {\int_{\bbR^N} d x\, \det\!J_t(x)\,e^{-S(z_t(x))}} 
\nonumber
\\
 &=\frac{
 \bigl\langle e^{i \theta_t(x)} 
 \mathcal{O}(z_t(x))
  \bigr\rangle_{S^{\rm eff}_t}}
 {\bigl\langle e^{i \theta_t(x)} 
 \bigr\rangle_{S^{\rm eff}_t}}. 
\label{eq:LT2}
\end{align}
Here, $S^{\rm eff}_t(x)$ and $\theta_t(x)$ are defined by 
\begin{align}
 e^{-S^{\rm eff}_t(x)} &\equiv e^{-{\rm Re}\,S(z_t(x))}\, |\det J_t(x)|,
\\
 e^{i \theta_t(x)} &\equiv e^{-i\,{\rm Im}\,S(z_t(x))}\, 
 e^{i\, {\rm arg}\, \det J_t(x)},
\end{align}
and $J_t(x)$ obeys the following differential equation 
\cite{Alexandru:2015xva} 
(see also footnote 2 of \cite{Fukuma:2017fjq}): 
\begin{align}
 \dot{J}_t &= [ H(z_t(x))\cdot J_t]^\ast,
 \quad
 J_{t=0} = \pmb{1} 
\label{eq:flow2}
\end{align}
with $H(z)\equiv (\partial_i \partial_j S(z))$. 
Under the flow \eqref{eq:flow1}, 
the action changes as $(d/dt) S(z_t(x))
 =\bigl|\partial_i S(z_t(x))\bigr|^2\geq 0$, 
and thus ${\rm Re}\,S(z_t(x))$ increases 
except at the critical points $z_\ast$ ($\partial_i S(z_\ast)=0$), 
while ${\rm Im}\,S(z_t(x))$ is kept constant. 
In particular, in the limit $t\to \infty$, 
the deformed region will approach a union of $N$-dimensional submanifolds
(Lefschetz thimbles) 
on each of which ${\rm Im}\,S(z)$ is constant, 
and thus the sign problem is expected to disappear there 
(except for a possible residual sign problem 
arising from the phase of the complex measure $dz$ 
and a possible global sign problem 
caused by phase cancellations among different thimbles). 
However, in the Monte Carlo calculation 
one cannot take the $t\to\infty$ limit na\"ively, 
because the potential barriers between different thimbles 
become infinitely high 
so that the whole configuration space cannot be explored sufficiently. 
This multimodality of distribution makes  
the Monte Carlo calculation impractical, 
especially when contributions from more than one thimble 
are relevant to estimating expectation values. 
A key proposal in \cite{Alexandru:2015sua}  
is to use a finite value of flow time 
that is large enough to avoid the sign problem 
but simultaneously is not too large 
so that the exploration in the configuration space is still possible. 
However, it is a difficult task to find such value of flow time 
in a systematic way, as we will discuss 
at the end of section \ref{sec:Hubbard} and in Appendix \ref{sec:fine-tuning}.

The TLTM \cite{Fukuma:2017fjq} is a tempering algorithm 
that uses the flow time as a tempering parameter. 
There, the global relaxation of the multimodal distribution is prompted 
by enabling configurations around different modes 
to easily communicate  
through transitions in ensembles at smaller flow times. 
Among other possible tempering algorithms, 
the parallel tempering algorithm \cite{Swendsen1986,Geyer1991} 
(also known as the replica exchange MCMC method; 
see \cite{Earl2005} for a review) 
is adopted in the TLTM \cite{Fukuma:2017fjq} 
because it does not need to introduce the probability weight factors 
of ensembles at various flow times 
and because most of relevant steps can be done in parallel processes.

In the TLTM 
(see Appendix \ref{sec:algorithm} 
for the summary of the algorithm),
we first fix the maximum flow time $T$ 
which should be sufficiently large 
such that the sign problem is reduced there. 
A possible criterion is 
that the sign average 
$|\langle e^{i \theta_{T}(x)} \rangle_{S^{\rm eff}_T}|$ 
is $O(1)$ in the absence of tempering. 
This process can be carried out by a test run with small statistics. 
We then enlarge the configuration space from $\bbR^N=\{x\}$ 
to the set of $A+1$ replicas, $(\bbR^N)^{A+1}=\{(x_0,x_1,\ldots,x_A)\}$. 
We assign to replicas $a$ ($a=0,1,\ldots,A$) 
the flow times $t_a$ with $t_0= 0 < t_1 < \cdots < t_A=T$. 
The action at replica $a$, $S^{\rm eff}_{t_a}(x_a)$, 
is obtained by solving \eqref{eq:flow1} and \eqref{eq:flow2} 
with its own initial conditions $z_{t=0}^i=x_a^i$, $J_{t=0}=\pmb{1}$. 
We set up an irreducible, aperiodic Markov chain 
for the enlarged configuration space 
such that the probability distribution for $\{(x_0,x_1,\ldots,x_A)\}$ 
eventually approaches the equilibrium distribution 
proportional to 
\begin{align}
 \prod_a \exp[-S^{\rm eff}_{t_a}(x_a)].
\label{global_distribution}
\end{align}
This can be realized 
by combining (a) the Metropolis algorithm 
(or the Hybrid Monte Carlo algorithm) 
in the $x$ direction at each fixed flow time 
and (b) the swap of configurations at two adjacent replicas. 
Each of the steps (a) and (b) can be done 
in parallel processes. 
After the system is well relaxed to global equilibrium, 
we estimate the expectation value at flow time $t_a$ 
[see \eqref{eq:LT2}]  
by using the subsample at replica $a$, 
$\{x_a^{(k)}\}_{k=1,2,\ldots,N_{\rm conf}}$, 
that is retrieved from the total sample 
$\{(x_0^{(k)},x_1^{(k)},\ldots,x_A^{(k)})\}_{k=1,2,\ldots,N_{\rm conf}}$: 
\begin{align}
 &\frac{
 \bigl\langle e^{i \theta_{t_a}(x)} 
 \mathcal{O}(z_{t_a}(x))
  \bigr\rangle_{S^{\rm eff}_{t_a}}}
 {\bigl\langle e^{i \theta_{t_a}(x)} 
 \bigr\rangle_{S^{\rm eff}_{t_a}}}
\nonumber
\\
 &\thickapprox
 \frac{\sum_{k=1}^{N_{\rm conf}}
 \exp[i \theta_{t_a}(x_a^{(k)})]\,
 \mathcal{O} (z_{t_a}(x_a^{(k)}))
  }
 {\sum_{k=1}^{N_{\rm conf}}
 \exp[i \theta_{t_a}(x_a^{(k)})] 
 }
 \equiv \bar{\mathcal{O}}_a.
\label{eq:estimate}
\end{align}
The original proposal in \cite{Fukuma:2017fjq} 
is to use \eqref{eq:estimate} at the maximum flow time, 
$\bar{\mathcal{O}}_{a=A}$, 
as an estimate of $\langle \mathcal{O} \rangle_S$.

Recall here that 
the left-hand side of \eqref{eq:estimate} 
is independent of $a$ due to Cauchy's theorem, 
and thus {\em the ratio $\bar{\mathcal{O}}_a$ 
should yield the same value 
within the statistical error margin 
if the system is well in global equilibrium}. 
In practice, this is not true for small $a$'s due to the sign problem, 
where 
the estimate of the sign average, 
$\bigl|\overline{e^{i\theta_{t_a}}}\bigr|\equiv
\bigl|(1/N_{\rm conf})\sum_k e^{i \theta_{t_a}(x^{(k)}_a)}\bigr|$, 
can be smaller than its statistical error 
($\simeq 1/\sqrt{2N_{\rm conf}}$; 
the value for the uniform distribution of phases). 
In this case, the statistical error of the ratio $\bar{\mathcal{O}}_a$ 
cannot be trusted, 
which means that such $\bar{\mathcal{O}}_a$ should not be used 
as an estimate of $\langle \mathcal{O} \rangle_S$. 

Based on the observation above, 
we now propose an algorithm 
which allows a precise estimation of $\langle \mathcal{O} \rangle_S$ 
with a criterion ensuring global equilibrium 
and the sufficiency of the sample size. 
First, we continue the sampling 
until we find some range of $a$ 
(to be denoted by $a=a_{\rm min},\ldots,a_{\rm max}(=A)$) 
in which 
$\bigl|\overline{e^{i\theta_{t_a}}}\bigr|$ 
are well above $1/\sqrt{2N_{\rm conf}}$ 
and $\bar{\mathcal{O}}_a$ take the same value 
within the statistical error margin.  
We will require that the $1 \sigma$ intervals 
around $\bigl|\overline{e^{i\theta_{t_a}}}\bigr|$ 
be above $3/\sqrt{2N_{\rm conf}}$. 
Then, we estimate $\langle \mathcal{O} \rangle_S$ 
by using the $\chi^2$ fit for
$\{\bar{\mathcal{O}}_a\}_{a=a_{\rm min},\ldots,a_{\rm max}}$ 
with a constant function of $a$. 
Global equilibrium and the sufficiency of the sample size 
are checked by looking at the optimized value of 
$\chi^2/{\rm DOF}=\chi^2/(a_{\max}-a_{\min})$. 
Note that the parameters determined by this procedure 
(such as $N_{\rm conf}$, $a_{\rm min}$, $a_{\rm max}=A$)
can vary depending on the choice of observable $\mathcal{O}$.

We close this section with a few comments. 
First, in the TLTM 
a sufficient overlap of the distributions at adjacent replicas 
is expected even for large flow times 
as long as the spacings are not too large. 
This is because the distributions at large $a$'s  
($\propto \exp[-S^{\rm eff}_{t_a}(x)]$) 
have peaks at the same points in $\bbR^N$  
that flow to critical points in $\bbC^N$. 
This is in sharp contrast with the situation in other tempered systems,
where the distribution often changes rapidly 
as a function of the tempering parameter 
so that an enough overlap 
cannot be achieved for realistically meaningful small spacings. 
Second, the optimal form of $t_a$ is a linear function of $a$ 
when flowed configurations are close to a critical point. 
This is because the optimal choice for the overall coefficients 
in tempering algorithms is exponential 
(see, e.g., \cite{Fukuma:2017wzs,Fukuma:2018qgv}) 
and because the real part of the action grows exponentially in flow time 
near critical points. 
Finally, the computational cost in the TLTM 
is expected to be $O(N^{3-4})$ 
due to the increase caused by the tempering algorithm 
(which will be $O(N^{0-1})$). 
Note that this growth of computational cost  
can be compensated by increasing the number of parallel processes.

\section{Application to the Hubbard model away from half-filling
\label{sec:Hubbard}}
Let $\Lambda=\{ {\bf x} \}$ be a $d$-dimensional lattice 
with $N_s$ lattice points. 
The Hubbard model describes nonrelativistic lattice fermions 
of spin one-half, 
and is defined by the Hamiltonian 
(including the chemical potential) 
\begin{align}
 H &= {}  -\kappa\,\sum_{\bx,\by}\sum_\sigma\,K_{\bx \by}\,c_{\bx,\sigma}^\dagger
 c_{\by,\sigma}
 - \mu\,\sum_\bx\,(n_{\bx,\uparrow} + n_{\bx,\downarrow} - 1)
\nonumber
\\
 & ~~~~ + U \sum_\bx\,(n_{\bx,\uparrow}-1/2)\,(n_{\bx,\downarrow}-1/2).
\label{Hubbard_H}
\end{align}
Here, $c_{\bx,\sigma}$ and $c_{\bx,\sigma}^\dagger$ 
are the annihilation and creation operators 
on site $\bx\in\Lambda$ with spin $\sigma$ $(=\uparrow,\downarrow)$ 
obeying the anticommutation relations 
$\{c_{\bx,\sigma},c_{\by,\tau}^\dagger\} 
 = \delta_{\bx \by}\,\delta_{\sigma \tau}$ and 
$\{c_{\bx,\sigma},c_{\by,\tau}\}
 = \{c_{\bx,\sigma}^\dagger,c_{\by,\tau}^\dagger\} = 0$, 
and $n_{\bx,\sigma}\equiv c_{\bx,\sigma}^\dagger c_{\bx,\sigma}$. 
$K_{\bx \by}$ is the adjacency matrix 
that takes a nonvanishing value ($\equiv 1$) only for nearest neighbors, 
and we assume the lattice to be bipartite.  
$\kappa\,(>0)$ is the hopping parameter, $\mu$ is the chemical potential, 
and $U\,(>0)$ represents the strength of the on-site repulsive potential. 
We have shifted $n_{\bx,\sigma}$ as 
$n_{\bx,\sigma}\to n_{\bx,\sigma}-1/2$ 
so that $\mu=0$ corresponds to the half-filling state, 
$\sum_\sigma\langle n_{\bx,\sigma}-1/2 \rangle = 0$.

We approximate the grand partition function ${\rm tr}\,e^{-\beta H}$ 
by using the Trotter decomposition with equal spacing $\epsilon$ 
($\beta = N_\tau \epsilon$), 
and rewrite it 
as a path integral over a Gaussian Hubbard-Stratonovich variable 
$\phi=(\phi_{\ell,\bx})$. 
Then the expectation value of the number density 
$n\equiv (1/N_s)\sum_{\bx}(n_{\bx,\uparrow}+n_{\bx,\downarrow}-1)$ 
is expressed as
(see Appendix \ref{sec:QMC} for the derivation)  
\begin{align}
 \langle n \rangle_S 
 &\equiv \frac{\int [d\phi]\,e^{-S[\phi]}\,n[\phi]}
 {\int [d\phi]\,e^{-S[\phi]}} 
 \quad
 \Bigl( [d\phi] \equiv \prod_{\ell,\bx} d\phi_{\ell,\bx} \Bigr),
\label{n_PI}
\\
 e^{-S[\phi]} &\equiv e^{-(1/2)\,\sum_{\ell,\bx} \phi_{\ell,\bx}^2}\,
 \det M^a[\phi]\,\det M^b[\phi],
\label{Hubbard_action1}
\\
 M^{a/b}[\phi] 
 &\equiv \pmb{1} + e^{\pm \beta \mu}\,\prod_\ell e^{\epsilon \kappa K}\, 
 e^{\pm \,i\sqrt{\epsilon U} \phi_\ell},
\label{Hubbard_action2}
\\
 n[\phi] &\equiv (i\sqrt{\epsilon U} N_s)^{-1}\,\sum_\bx \phi_{\ell=0,\bx},
\label{Hubbard_action3}
\end{align}
where $\phi_\ell\equiv(\phi_{\ell,\bx}\,\delta_{\bx\by})$ 
and $\prod_\ell$ is a product in descending order. 
Note that $n[\phi]$ in \eqref{Hubbard_action3} can be replaced by 
$(i\sqrt{\epsilon U} N_\tau N_s)^{-1}\,\sum_{\ell,\bx} \phi_{\ell,\bx}$, 
which is more preferable in Monte Carlo calculations 
because statistical errors will be reduced 
due to the averaging over $\ell$. 
The charge-charge correlation, 
$\langle n_\bx\,n_\by \rangle_S$ 
($n_\bx \equiv n_{\bx,\uparrow}+n_{\bx,\downarrow}-1$), 
can also be evaluated as a path integral 
by simply replacing $n_\bx$ by $(i\sqrt{\epsilon U})^{-1}\,\phi_{\ell=0,\bx}$ 
when $\bx \neq \by$.
As for the observables that are not directly constructed from $n_\bx$, 
the expectation values can be evaluated by using the formula \eqref{BdD}.

We now apply the TLTM to the Hubbard model 
on a two-dimensional periodic square lattice of size $2\times 2$ 
(thus $N_s=4$) 
with $N_\tau=5$. 
We first estimate $\langle n \rangle_S$ numerically 
by using the expressions \eqref{n_PI}--\eqref{Hubbard_action3} 
for various values of $\beta\mu$ 
with other parameters fixed to be $\beta\kappa=3$, $\beta U=13$. 
Note that the physical quantities depend only on the dimensionless parameters 
$\beta\mu$, $\beta\kappa$, $\beta U$ for fixed $N_\tau$.

\begin{figure}[ht]
\includegraphics[width=42.5mm]{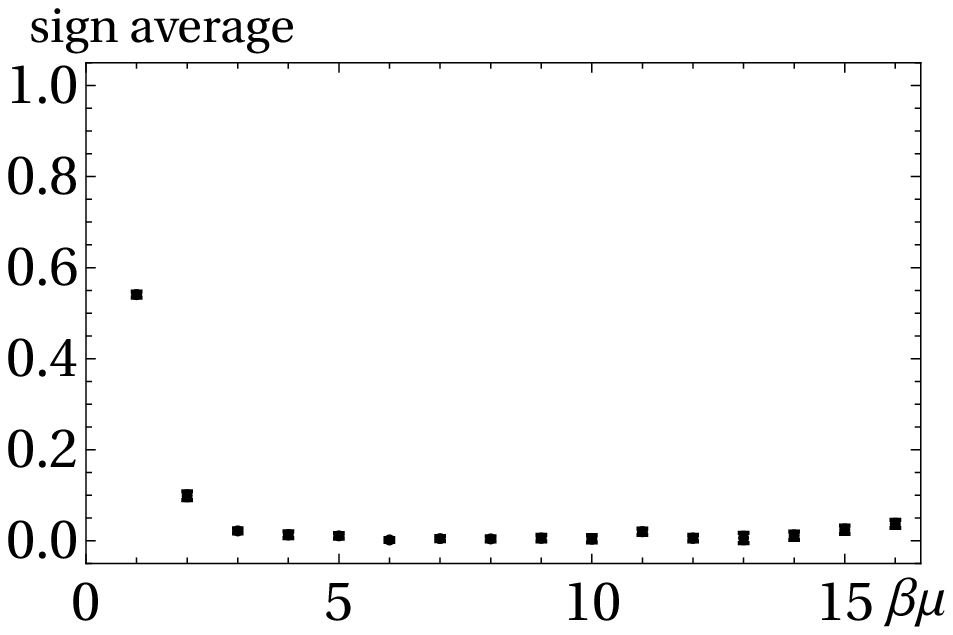}%
\includegraphics[width=42.5mm]{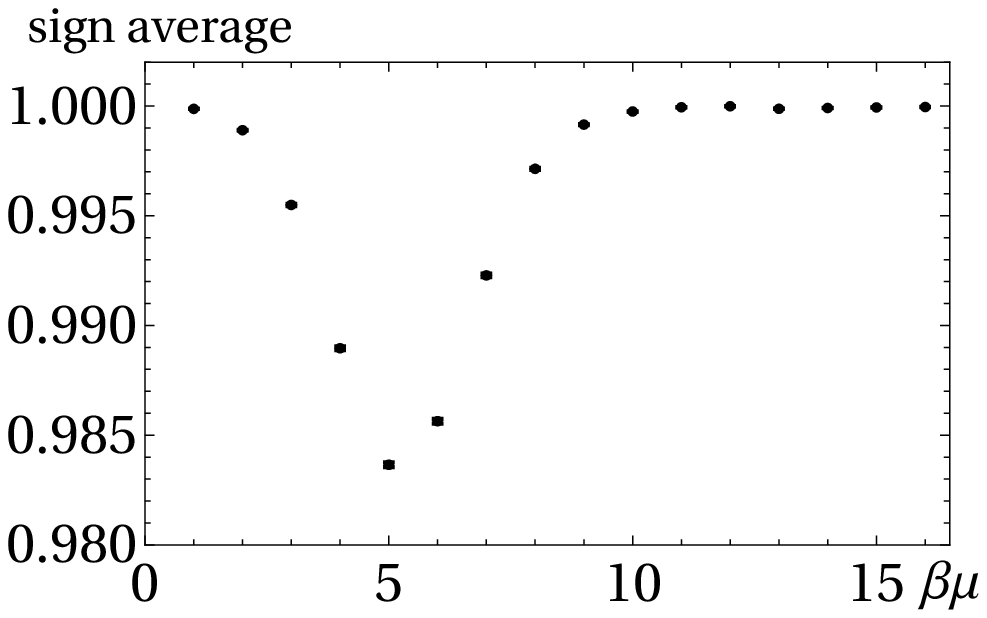} 
\caption{
\label{fig:reweighting}
 (Left) the sign averages obtained by the reweighting method (flow time $T=0$) 
 for the complex action \eqref{Hubbard_action1} 
 with the Gaussian Hubbard-Stratonovich variable. 
 (Right) the sign averages obtained by using ALF 
 with the $M_z$ parametrization.
}
\end{figure}

The complex action \eqref{Hubbard_action1} gives rise to 
a serious sign problem, 
as can be seen in the left panel of Fig.~\ref{fig:reweighting}. 
However, we should note that 
the extent of the seriousness of the sign problem 
heavily depends on the choice of the Hubbard-Stratonovich variables, 
and actually, the sign problem can be avoided 
for the above parameters 
within the BSS (Blankenbecler, Scalapino and Sugar)-QMC method 
\cite{Blankenbecler:1981jt}. 
In fact, the right panel of Fig.~\ref{fig:reweighting} 
shows the sign averages calculated by using a public code 
called ALF (Algorithms for Lattice Fermions) \cite{Bercx:2017pit} 
that is based on the discrete variables 
introduced in \cite{Motome:1997,Assaad:1997}. 
We see that the sign averages are above 0.98 
for all the range of $\beta\mu$ studied here.%
\footnote{ 
  We thank a referee for suggesting us to investigate this point. 
} 

Following the general prescription 
and writing $x=(x^i)=(\phi_{\ell,\bx})$ $(i=1,\ldots,N)$ with $N=N_\tau N_s$, 
we introduce the enlarged configuration space 
$(\mathbb{R}^N)^{A+1}=\{(x_0,x_1,\ldots,x_A)\}$. 
We here brief the setup of the parameters relevant to the TLTM 
(see Appendix \ref{sec:parameters} for more details). 
We set $t_a$ to be piecewise linear in $a$ 
with a single breakpoint 
whose position will be tuned 
such that the acceptance rates of the swapping process 
at adjacent replicas are almost the same for all pairs 
(being roughly above 40\%).%
\footnote{ 
 This functional form of $t_a$ is best suited to the case 
 where the deformed region reaches the vicinity 
 of all the relevant Lefschetz thimbles at almost the same flow time 
 and such a linear form is effective also for the transient period. 
} 
For each value of $\beta\mu$, 
we make a test run with small statistics 
to adjust parameters. 
This gives the values $T/(\beta\mu)=1/12$--$1/10$, $A=8$--$12$, 
$N_{\rm conf}=5,000$--$25,000$, 
varying on the value of $\beta\mu$. 
We make a sampling after discarding 5,000 configurations, 
and from the obtained data 
$\{\bar{n}_a\}_{a=a_{\rm min},\ldots,a_{\rm max}}$ 
we estimate $\langle n \rangle_S$ 
by using the $\chi^2$ fit.

\begin{figure}[ht]
\includegraphics[width=42.5mm]{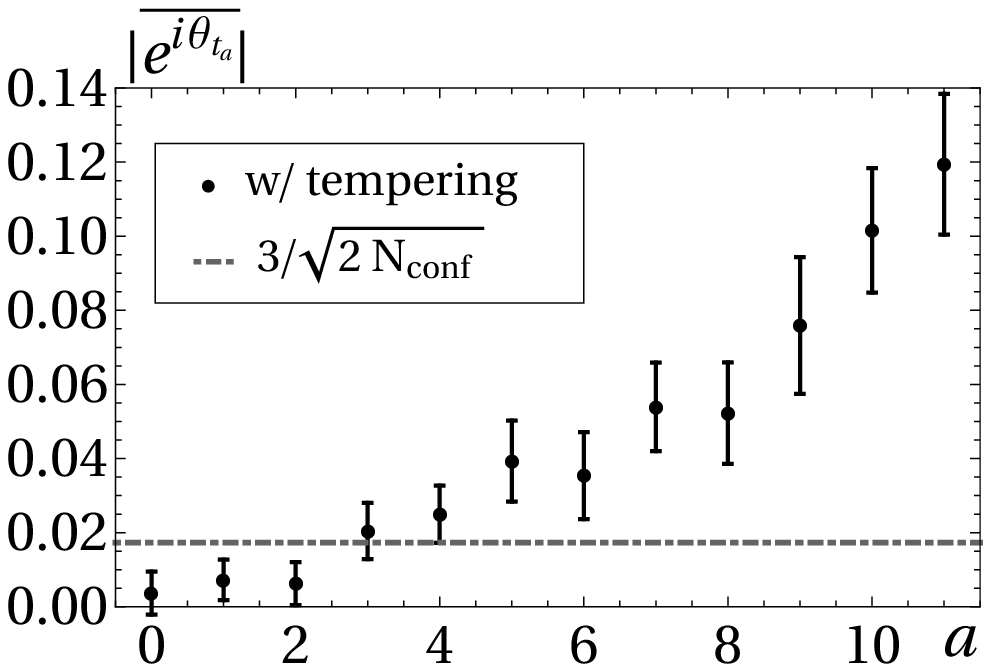} 
\includegraphics[width=42.5mm]{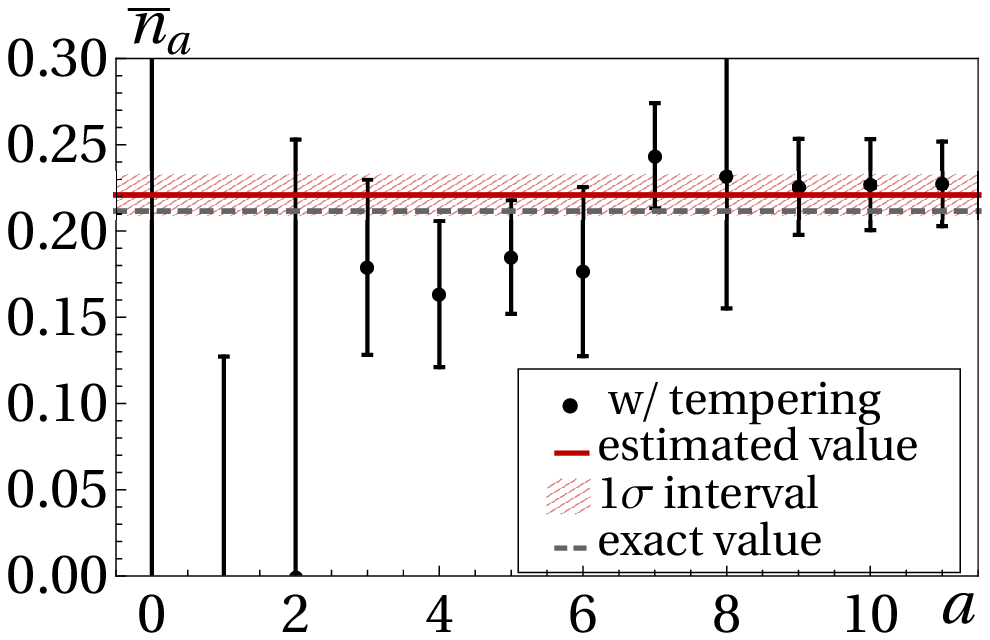}%
\caption{
\label{fig:mu5_tltm}
 With tempering ($\beta\mu=5$). 
 (Left) the sign averages 
 at various replicas. 
 The horizontal dashed line represents $3/\sqrt{2N_{\rm conf}}= 0.017$. 
 (Right) the data $\bar{n}_a$. 
 The solid red line with a shaded band represents 
 the estimate of $\langle n \rangle_S$ with $1 \sigma$ interval. 
 The gray dashed line represents the exact value.
}
\end{figure}

As an example, 
let us see Fig.~\ref{fig:mu5_tltm}, 
which shows
$\bigl|\overline{e^{i\theta_{t_a}}}\bigr|$ 
and $\bar{n}_a$ at various replicas for $\beta\mu=5$. 
The left panel shows that the $1\sigma$ intervals 
around  $\bigl|\overline{e^{i\theta_{t_a}}}\bigr|$ 
are above $3/\sqrt{2 N_{\rm conf}}$ 
for $a=5,\ldots,11$ 
(and thus we set $a_{\rm min}=5$ and $a_{\rm max}=11$). 
The right panel shows that 
the data $\{\bar{n}_a\}$ in this range give the same value
within the statistical error margin. 
The $\chi^2$ fit gives the estimate 
$\langle n\rangle_S \approx 0.221 \pm 0.012$ 
(exact value: 0.212)
with $\chi^2/{\rm DOF}=0.45$.

\begin{figure}[ht]
\includegraphics[width=86mm]{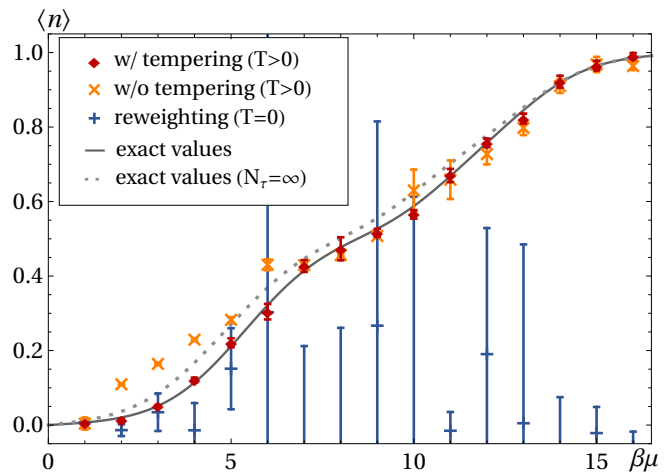}%
\caption{
\label{fig:n-mu}
 The expectation values of the number density operator, 
 $\langle n \rangle_S$ $(N_\tau=5)$. 
 The results obtained with tempering 
 correctly reproduce the exact values. 
 The exact values for $N_\tau=\infty$ are also displayed for comparison. 
}
\end{figure}

\begin{figure}[ht]
\includegraphics[width=42.5mm]{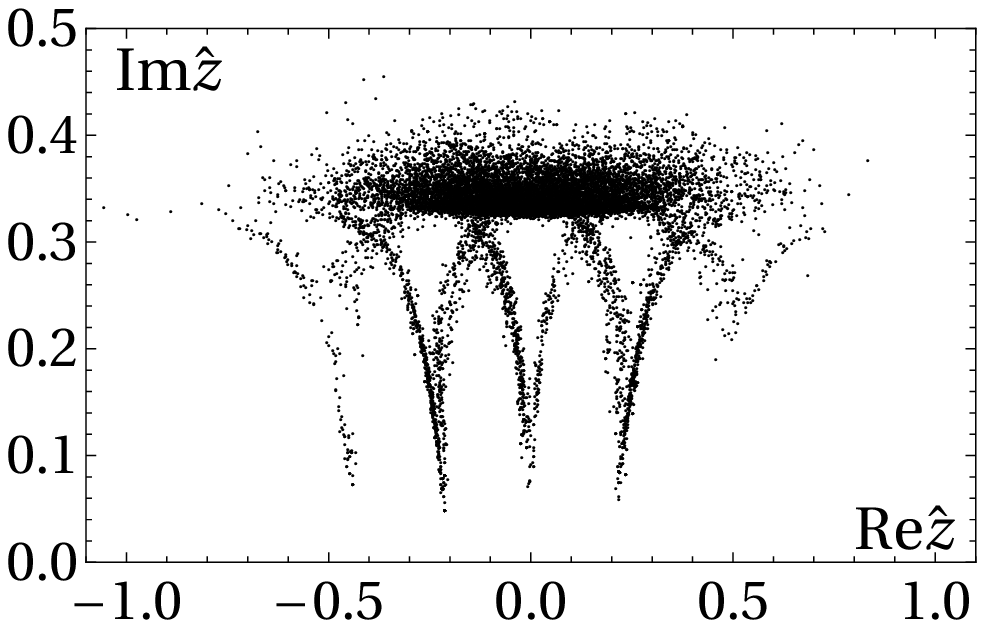} 
\includegraphics[width=42.5mm]{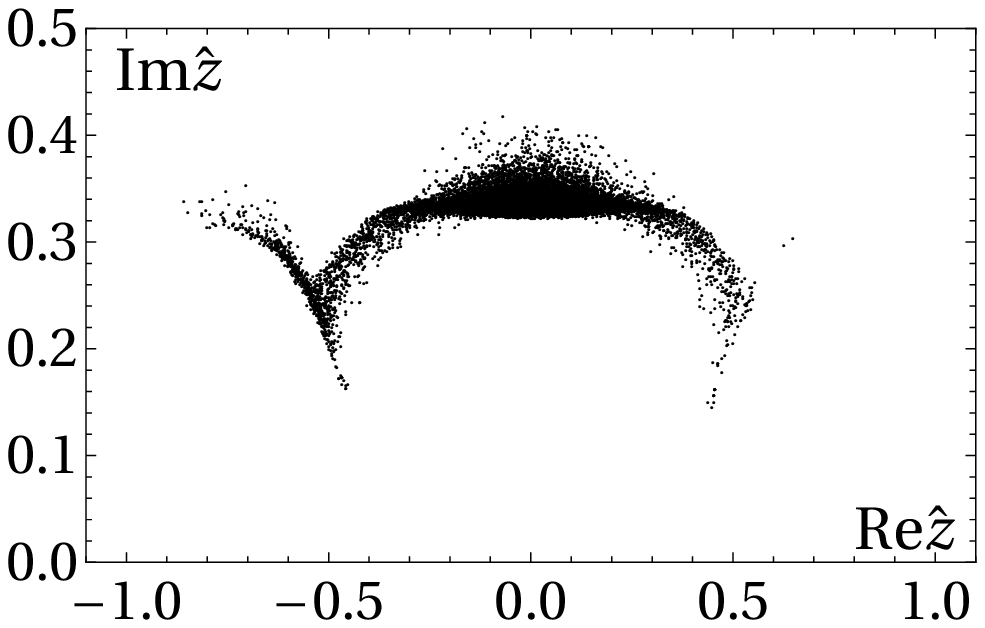}%
\caption{
\label{fig:mu5_a11}
 The distribution of $\hat{z}$. 
 (Left) with tempering. (Right) without tempering.  
}
\end{figure}

Figure \ref{fig:n-mu} shows the thus-obtained numerical estimates 
of $\langle n \rangle_S$ 
as a function of $\beta\mu$. 
We also display the estimates 
obtained without tempering (at the same maximum flow times $T$) 
and those from the original reweighting method (i.e.\ $T=0$),
together with the values 
obtained by the explicit evaluation of the trace 
under the Trotter decomposition
with $N_\tau=5$ and for the continuum imaginary time 
(i.e.\ $N_\tau=\infty$)
(see Appendix \ref{sec:Trotter}).
We see that 
the exact values are correctly reproduced 
when the tempering is implemented, 
while there are significant deviations when not implemented. 
As in the $(0+1)$-dimensional massive Thirring model \cite{Fukuma:2017fjq}, 
the deviation reflects the fact that 
the relevant thimbles are not sampled sufficiently. 
In fact, from Fig.~\ref{fig:mu5_a11}, 
which shows the distribution 
of averaged flowed configurations $\hat{z}\equiv (1/N)\sum_i z_T^i$
at $T=0.5$ for $\beta\mu=5$,
we see that, 
although the flowed configurations are widely distributed 
over many thimbles when the tempering is implemented, 
they are restricted to only a small number of thimbles 
when not implemented.

\begin{figure}[ht]
\includegraphics[width=86mm]{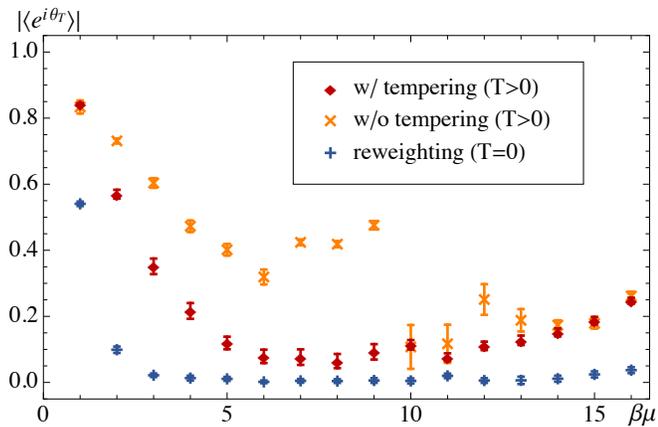}%
\caption{
\label{fig:denom-mu}
 The sign averages at $T$, 
 $|\langle e^{i\theta_T(x)}\rangle_{S^{\rm eff}_T}|$
}
\end{figure}

Three comments are in order. 
First, a larger value of the sign average does not necessarily mean 
a better resolution of the sign problem, 
as can be seen from Fig.~\ref{fig:denom-mu}. 
In fact, when only a very few thimbles are sampled, 
the sign average can become larger than the value in the correct sampling 
due to the absence of phase mixtures among different thimbles. 

\begin{figure}[ht]
\includegraphics[width=86mm]{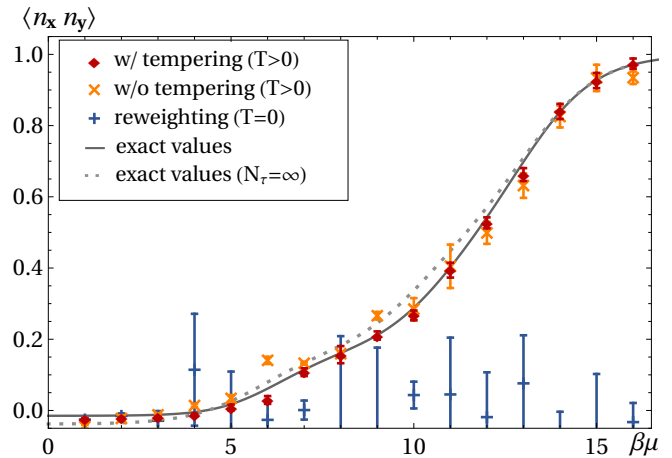}%
\caption{
\label{fig:nn-mu}
 The nearest-neighbor charge-charge correlations 
 $\langle n_\bx n_\by \rangle_S$ ($N_\tau=5$)
}
\end{figure}

Second, 
whether the multimodality can affect the estimates of expectation values 
depends on the choice of observables. 
In fact, from the discrepancies of the sign averages 
in Fig.~\ref{fig:denom-mu}, 
we see that the multimodality must be severe 
in the region $\beta\mu\leq 9$. 
However, the estimates of $\langle n \rangle_S$ almost agree 
between the two methods with and without tempering 
in the range $7\leq \beta\mu \leq 9$. 
This means that the operator $n$ is not sensitive to the multimodality 
in this range. 
To find an observable that is sensitive to the multimodality, 
we estimated the nearest-neighbor charge-charge correlation 
$\langle n_\bx \, n_\by \rangle_S$ 
with the same sample.%
\footnote{
 We thank the referee for suggesting us 
 to investigate the expectation values of observables 
 other than the number density operator. 
} 
The results are shown in Fig.~\ref{fig:nn-mu}, 
where we see a significant discrepancy at $\beta\mu=9$ 
between the two methods. 

\begin{figure}[ht]
\includegraphics[width=86mm]{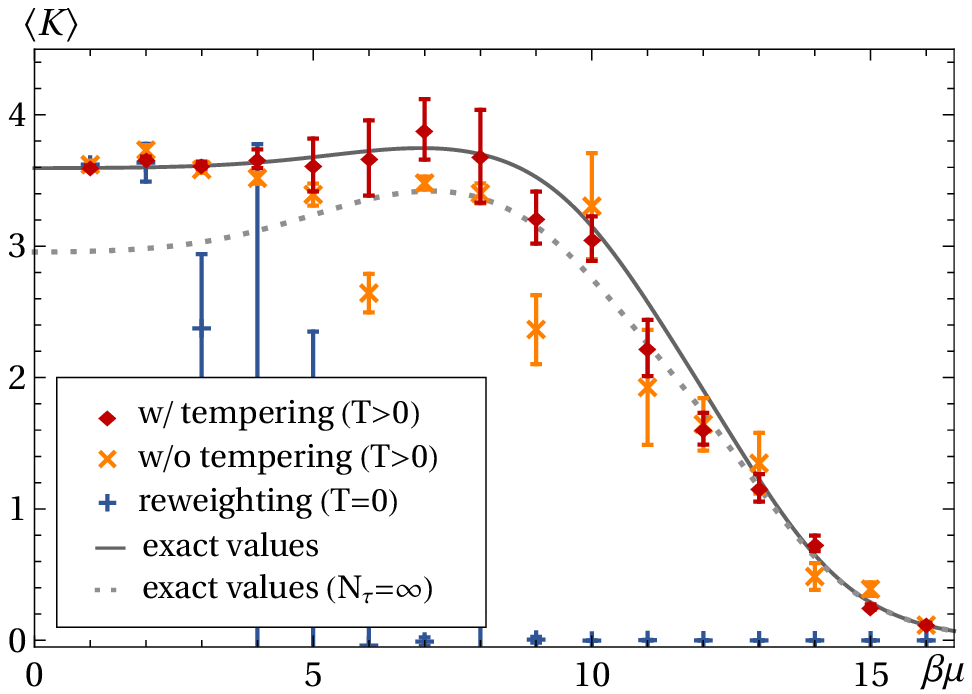}%
\caption{
\label{fig:K-a}
 The kinetic energies $\langle K \rangle_S$ ($N_\tau=5$)
}
\end{figure}

Such discrepancies become more manifest 
if we look at the observables 
that are not directly constructed from the number density operator $n_\bx$. 
As an example, 
we show in Fig.~\ref{fig:K-a} 
the expectation values of the kinetic energy operator 
(without the factor ``$-\kappa$'')
$K\equiv
\sum_{\bx,\by}\sum_\sigma K_{\bx \by}\, c^\dag_{\bx,\sigma}c_{\by,\sigma}$, 
which are estimated for the same sample as above 
by using the formula \eqref{BdD}. 
We there notice two things. 
One is that 
the discrepancies between the two methods 
now become significant 
for all the range $7\leq \beta\mu \leq 9$.  
The other is that the precision of the TLTM 
becomes worse compared to the case for the observables 
that are constructed solely from $n_\bx$. 
In fact, those observables 
that are \textit{not} directly constructed from $n_\bx$ 
(such as $K$)
contain matrix elements of $M^{a/b}[\phi]^{-1}$, 
and may have divergently large values 
in the vicinity of zeros of the fermion determinants $\det M^{a/b}[\phi]$. 
In this case, 
precise estimation will require a larger sample size 
and a more accuracy in integrating flow equations 
compared with operators constructed solely from $n_\bx$.
We expect that a similar attention must be paid 
when one applies the TLTM to finite density QCD. 
We leave a further investigation of this point as a future investigation. 

\begin{figure}[ht]
\includegraphics[width=42.5mm]{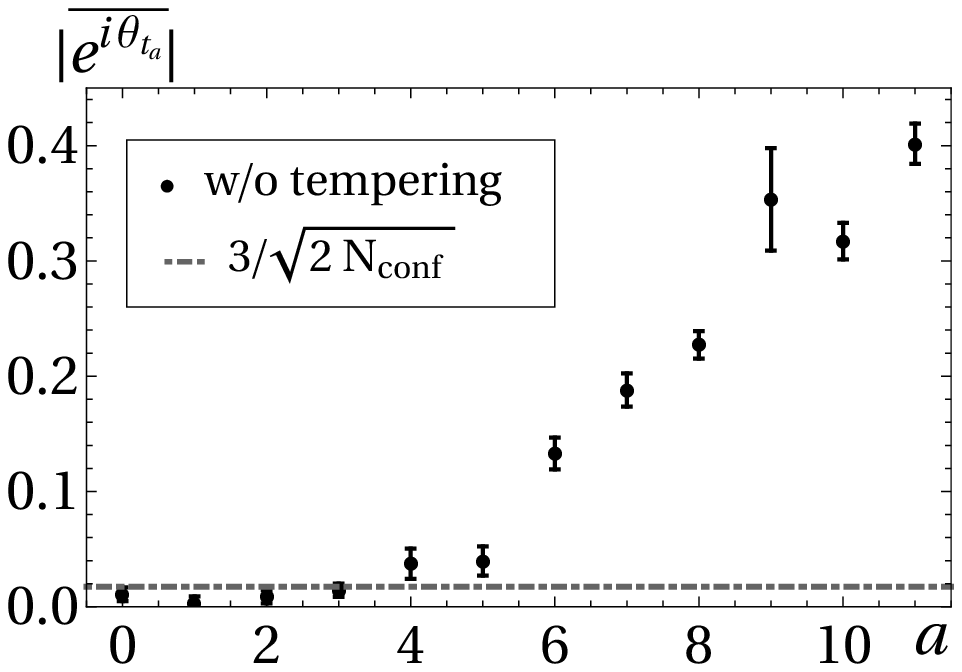} 
\includegraphics[width=42.5mm]{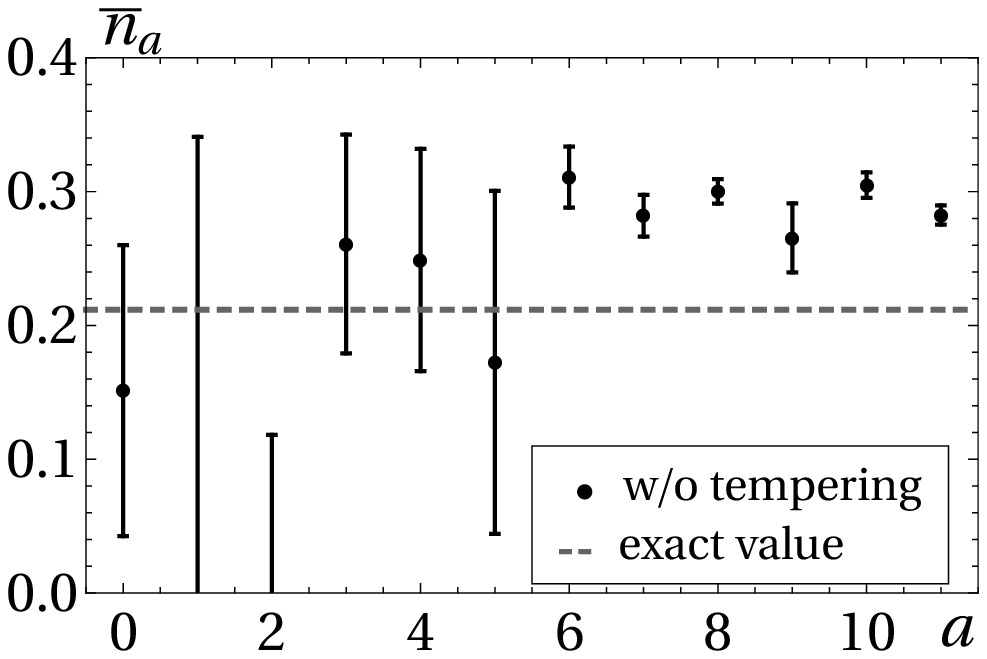}%
\caption{
\label{fig:mu5_gltm}
 Without tempering ($\beta\mu=5$). 
 (Left) the sign averages. 
 (Right) the estimates $\bar{n}_a$. 
 There is no such flow time that clearly avoids 
 both the sign and multimodal problems simultaneously 
 (at least for the present spacings).
}
\end{figure}

Finally, 
from Fig.~\ref{fig:mu5_gltm}, we see that
it should be a difficult task 
to find an intermediate flow time (without tempering) 
that avoids both the sign problem (severe at smaller flow times) 
and the multimodal problem (severe at larger flow times) 
(see Appendix \ref{sec:fine-tuning} for more detailed discussions).
Generically, 
flowed configurations repeatedly experience the event 
at which they get trapped to fewer number of Lefschetz thimbles, 
so that there is a large ambiguity 
in distinguishing the larger and smaller flow times 
in the first place.

\section{Conclusion and outlook
\label{sec:conclusion}}
In this paper, 
we proposed an algorithm for the TLTM 
which allows a precise estimation of expectation values. 
We confirm the effectiveness 
by applying it to the two-dimensional Hubbard model 
away from half-filling.

We should stress that our study in this paper 
is still at an exploratory level. 
In fact, the lattice must be enlarged much more 
both in the spatial and imaginary time directions 
to claim the validity of our method 
for the sign problem in the Hubbard model, 
revealing the phase structure of the model. 
In doing this, 
it should be important to check 
whether the computational scaling is actually $O(N^{3-4})$ 
as expected.

More generally, we should keep developing the algorithm further 
so that it can be more easily applied 
to the three major problems listed in Introduction. 
There should also be other interesting branches of fields 
where the TLTM may shed new light on the theoretical understanding 
through a numerical analysis, 
such as the Chern-Simons theory \cite{Witten:2010cx} 
and matrix models that generate random volumes 
\cite{Fukuma:2015xja}.

When we were preparing the first version of the manuscript, 
there appeared an interesting paper 
\cite{Ulybyshev:2019a} 
(see also its detailed version \cite{Ulybyshev:2019b}), 
where the sign and ergodicity problems are also studied 
for the Lefschetz thimble method 
applied to the Hubbard model away from half-filling. 
In our method (TLTM), 
the two problems are solved simultaneously 
by tempering the system with the flow time, 
where one does not need to know a detailed structure of thimbles. 
In contrast, 
in \cite{Ulybyshev:2019a,Ulybyshev:2019b} 
they redundantly introduce 
two continuous Gaussian Hubbard-Stratonovich variables 
with a parameter representing the mixture of the two variables 
(see also \cite{Ulybyshev:2017hbs}). 
With knowledge of thimble structures, 
they tune the parameter in such a way that 
only a few number of thimbles become relevant to the evaluation, 
and obtain results 
for a $2\times 2$ hexagonal lattice $(N_s=8)$ 
with $N_\tau = 384$ and $\beta=30$. 
It would be interesting 
to introduce such redundant integration variables also in the TLTM 
so as to reduce the global sign problem 
(possible cancellation of phases among different thimbles), 
which we observe also depends heavily 
on the choice of integration variables. 

\begin{acknowledgments}
The authors thank Yoshimasa Hidaka, Issaku Kanamori, Norio Kawakami, 
Yoshio Kikukawa, Jun Nishimura, Akira Ohnishi, Masaki Tezuka, Asato Tsuchiya 
and Urs Wenger
for useful discussions. 
They also thank an anonymous referee of Physical Review D 
for giving us valuable comments, 
which were very helpful in improving the first version of the manuscript. 
This work was partially supported by JSPS KAKENHI 
(Grant Numbers 16K05321, 18J22698 and 17J08709) 
and by SPIRITS 2019 of Kyoto University (PI: M.F.).
\end{acknowledgments}


\appendix


\section{Summary of the algorithm
\label{sec:algorithm}}
We summarize the algorithm of the TLTM 
(we partially repeat the presentation of \cite{Fukuma:2017fjq}): 

\noindent
$\bullet$~{\underline{Step 0}.}
We fix the maximum flow time $T$ 
which should be sufficiently large 
such that the sign problem is reduced there. 
A possible criterion is that 
the sign average $|\langle e^{i \theta_{T}(x)} \rangle_{S^{\rm eff}_T}|$ 
is $O(1)$ in the absence of tempering. 
This can be carried out by a test run with small statistics. 
We then pick up flow times $\{t_a\}$ from the interval $[0,T]$ 
with $t_0=0 < t_1 < \cdots < t_A = T$. 
The values of $A$ and $t_a$ are determined manually or adaptively
to optimize the acceptance rate in Step 3 below. 
Practically, once $A$ is determined, 
$t_a$ can be chosen to be a piecewise linear function of $a$ 
[see the argument for \eqref{piecewise_linear}].

\noindent
$\bullet$~{\underline{Step 1}.}
For each replica $a$, 
we choose an initial value $x_a \in \bbR^N$ 
and numerically solve the differential equations 
\eqref{eq:flow1} and \eqref{eq:flow2} 
to obtain the triplet 
$(x_a,z_a\equiv z_{t_a}(x_a),
J_a\equiv J_{t_a}(x_a))$.

\noindent
$\bullet$~{\underline{Step 2}.}
For each replica $a$, 
we use the Metropolis algorithm to update the value of $x_a$. 
To be explicit, we take a value $x'_a$ from $x_a$ 
using a symmetric proposal distribution, 
and recalculate the triplet $(x'_a,z'_a,J'_a)$ 
using the $x'_a$ as the initial value. 
We then update $x_a$ to $x'_a$ 
with the probability ${\rm min}(1,e^{-\Delta S_a})$, 
where
\begin{align}
 \Delta S_a
 &\equiv S^{\rm eff}_{t_a}(x'_a)-S^{\rm eff}_{t_a}(x_a)
\nonumber
\\
 &=({\rm Re}\,S(z'_a)-\ln\,\bigl|\det\!J'_a\bigr|)
 -({\rm Re}\,S(z_a)-\ln\,\bigl|\det\!J_a\bigr|).
\end{align}
We repeat the process sufficiently many times 
such that local equilibrium is realized for each $a$. 
Step 1 and Step 2 can be performed in parallel processes. 

\noindent
$\bullet$~{\underline{Step 3}.}
We swap the configurations at two adjacent replicas $a$ and $a+1$ 
by updating $(x_a,x_{a+1})=(x,y)$ to $(x'_a,x'_{a+1})=(y,x)$ with the probability 
\begin{align}
 w_a(x,y)={\rm min}\Bigl(1,\,
 e^{-S^{\rm eff}_{t_a}(y)-S^{\rm eff}_{t_{a+1}}(x)
 +S^{\rm eff}_{t_a}(x)+S^{\rm eff}_{t_{a+1}}(y)}\Bigr). 
\end{align}
One can easily see that this satisfies the detailed balance condition 
with respect to the global equilibrium distribution \eqref{global_distribution} 
because
\begin{align}
 w_a(x,y)\,e^{-S^{\rm eff}_{t_a}(x)-S^{\rm eff}_{t_{a+1}}(y)}
 = w_a(y,x)\,e^{-S^{\rm eff}_{t_a}(y)-S^{\rm eff}_{t_{a+1}}(x)}.
\end{align}
We repeat the process several times 
so as to reduce autocorrelations. 
This procedure can also be performed in parallel processes 
by choosing a set of independent pairs.

\noindent
$\bullet$~{\underline{Step 4}.}
By repeating Step 2 and Step 3, 
we obtain a sequence of triplets,
\begin{align}
 \{(x_a^{(k)},z_a^{(k)},J_a^{(k)})\}_{k=1,2,\ldots,N_{\rm conf}},
\end{align}
for each $a$, 
with which we estimate the expectation value at flow time $t_a$:
\begin{align}
 \frac{
 \bigl\langle e^{i \theta_{t_a}(x)} 
 \mathcal{O}(z_{t_a}(x))
  \bigr\rangle_{S^{\rm eff}_{t_a}}}
 {\bigl\langle e^{i \theta_{t_a}(x)} 
 \bigr\rangle_{S^{\rm eff}_{t_a}}}
 &\thickapprox
 \frac{\sum_{k=1}^{N_{\rm conf}}
 e^{i \theta^{(k)}_a} 
 \mathcal{O}\bigl(z_a^{(k)}\bigr)
  }
 {\sum_{k=1}^{N_{\rm conf}}
 e^{i \theta^{(k)}_a} 
 }
 \equiv \bar{\mathcal{O}}_a
\nonumber
\\
 &\qquad [ \theta_a^{(k)} \equiv \theta_{t_a}(x_a^{(k)}) ].
\label{eq:sum}
\end{align}
Here, $N_{\rm conf}$ is chosen to be large enough 
so that we find some range of $a$ 
(to be denoted by $a=a_{\rm min},\ldots,a_{\rm max}$ with $a_{\rm max}=A$) 
in which the $1 \sigma$ intervals 
around 
$\bigl|\overline{e^{i\theta_{t_a}}}\bigr|
=\bigl|(1/N_{\rm conf})\sum_k e^{i \theta_{t_a}(x^{(k)}_a)}\bigr|$ 
are above $3/\sqrt{2N_{\rm conf}}$ 
and $\bar{\mathcal{O}}_a$ 
take the same value within the statistical error margin.  

\noindent
$\bullet$~{\underline{Step 5}.}
The expectation value of $\langle \mathcal{O} \rangle_S$ 
is estimated by the $\chi^2$ fit 
from the data $\{\bar{\mathcal{O}}_a\}_{a=a_{\rm min},\ldots,a_{\rm max}}$ 
with a constant function of $a$. 
Global equilibrium and the sufficiency of the sample size $N_{\rm conf}$ 
is checked by looking at the optimized value of 
$\chi^2/{\rm DOF}=\chi^2/(a_{\max}-a_{\min})$.

In the above algorithm, 
we have implicitly assumed 
that the action at $t_0=0$ does not exhibit multimodality. 
If this is not the case, 
we further introduce other parameters 
(such as the overall coefficient of the action) as extra tempering parameters 
or prepare flow times $\{t_a\}$ with $t_0 < 0$ \cite{Fukuma:2017fjq}.

\section{Derivation of eqs.\ \eqref{n_PI}--\eqref{Hubbard_action3}
\label{sec:QMC}}
For a bipartite lattice, 
we specify which sublattice $\bx$ belongs to by the sign $(-1)^\bx=\pm 1$.  
We first make the so-called particle-hole transformation, 
$c_{\bx,\uparrow}=a_\bx$ and $c_{\bx,\downarrow}=(-1)^\bx\,b_\bx^\dagger$. 
Then the one-body part $H_1$ and 
the two-body part $H_2$ of the Hamiltonian \eqref{Hubbard_H} 
are rewritten, respectively, as 
\begin{align}
 H_1 
 &={} -\sum_{\bx,\by}\, (\kappa K + \mu\, \pmb{1})_{\bx\by}\,
 a_\bx^\dagger a_\by
 -\sum_{\bx,\by}\, (\kappa K - \mu\, \pmb{1})_{\bx\by}\,b_\bx^\dagger b_\by,
\label{H1}
\\
 H_2 &={} -U\,\sum_\bx (n^a_\bx-1/2)(n^b_\bx-1/2)
\nonumber
\\
 &= (U/2)\,\sum_\bx\,(n^a_\bx-n^b_\bx)^2 - N_s U/4.
\label{H2}
\end{align}
In the last equation, we have used the identity 
$n^a_\bx\,(\equiv a_\bx^\dagger a_\bx) = (n^a_\bx)^2$ and 
$n^b_\bx\,(\equiv b_\bx^\dagger b_\bx) = (n^b_\bx)^2$. 
Note that the number density operator is written as 
\begin{align}
 n \equiv (1/N_s)\,\sum_\bx\,(n_{\bx,\uparrow}+n_{\bx,\downarrow}-1)
 = (1/N_s)\,\sum_\bx\,(n^a_\bx - n^b_\bx).
\label{ndensity} 
\end{align}

In order to perform a Monte Carlo simulation, 
we approximate $e^{-\beta H}$ in the grand partition function 
by using the Trotter decomposition 
with equal spacing $\epsilon$ ($\beta=N_\tau \epsilon$):
\begin{align}
 e^{-\beta H} = (e^{-\epsilon (H_1+H_2)})^{N_\tau}
 \simeq (e^{-\epsilon H_1}\,e^{-\epsilon H_2})^{N_\tau},
\end{align}
and rewrite $e^{-\epsilon H_2}$ at the $\ell$-th position from the right 
to the exponential of a fermion bilinear 
by using a Gaussian Hubbard-Stratonovich variable $\phi_{\ell,\bx}$:
\begin{align}
 e^{-\epsilon H_2}
 &= e^{N_s \epsilon U/4}\,e^{-(\epsilon U/2)\,\sum_\bx(n^a_\bx-n^b_\bx)^2}
\nonumber
\\
 &= e^{N_s \epsilon U/4}\,\prod_\bx \int \frac{d\phi_{\ell,\bx}}{\sqrt{2\pi}}\,
 e^{-(1/2)\,\phi_{\ell,\bx}^2 
 + \,i\sqrt{\epsilon U}\,\phi_{\ell,\bx} (n^a_\bx-n^b_\bx) }. 
\end{align}
Then, the approximated grand partition function takes 
the following path integral form: 
\begin{align}
 Z_{\rm QMC}
 &\equiv {\rm tr}\,\bigl[(e^{-\epsilon H_1}\,e^{-\epsilon H_2})^{N_\tau}\bigr]
\nonumber
\\
 &= (e^{\epsilon U/4}/\sqrt{2\pi})^{N_\tau N_s}
 \int [d\phi]\,e^{-S[\phi]}.
\end{align}
Here, $[d\phi] \equiv \prod_{\ell,\bx} d\phi_{\ell,\bx}$, 
and the action $S[\phi]$ is given by 
\begin{widetext}
\begin{align}
 e^{-S[\phi]} 
 &= e^{-\sum_{\ell,\bx} (1/2)\,\phi_{\ell,\bx}^2}\,
 {\rm tr}_a\,\prod_\ell\,
 e^{\epsilon\sum_{\bx,\by}(\kappa K + \mu\, \pmb{1})_{\bx\by}\,
 a_\bx^\dagger a_\by}\,
 e^{\sum_\bx\,(i\sqrt{\epsilon U}\,\phi_{\ell,\bx})\,a_\bx^\dagger a_\bx}
\nonumber
\\
 &~~~~~\times 
 {\rm tr}_b\,\prod_\ell\,
 e^{\epsilon\sum_{\bx,\by}(\kappa K - \mu\, \pmb{1})_{\bx\by}\,
 b_\bx^\dagger b_\by}\,
 e^{\sum_\bx\,(-i\sqrt{\epsilon U}\,\phi_{\ell,\bx})\,b_\bx^\dagger b_\bx},
\label{action_fermion}
\end{align}
\end{widetext}
where $\prod_\ell$ is an ordered product 
($\prod_\ell f_\ell \equiv f_{N_\tau-1}\cdots f_1 f_0$), 
and ${\rm tr}_a$ (or ${\rm tr}_b$) represents the trace 
over the Fock space created by $a_\bx^\dagger$ (or by $b_\bx^\dagger$). 
The fermion trace in \eqref{action_fermion} can be evaluated explicitly 
by using the following formulas 
that hold for the operator 
$\hat{A}\equiv\sum_{\bx,\by}\,A_{\bx\by}\,a_\bx^\dagger a_\by$ 
constructed from an $N_s\times N_s$ matrix $A=(A_{\bx\by})$: 
\begin{align}
 &e^A\,e^B = e^C ~~\Rightarrow~~ e^{\hat{A}}\,e^{\hat{B}} = e^{\hat{C}},
\\
 &{\rm tr}\,e^{\hat{A}} = \det\, (\pmb{1}+e^A).
\end{align}
(The first equation can be readily proved 
by the fact that $A\mapsto\hat{A}$ is a Lie algebra homomorphism. 
The second equation can be easily understood 
by moving to a diagonalizing basis for $A$.) 
We thus find that the action becomes 
\begin{align}
 e^{-S[\phi]} &= e^{-(1/2)\,\sum_{\ell,\bx} \phi_{\ell,\bx}^2}\,
 \det M^a[\phi]\,\det M^b[\phi],
\\
 M^{a/b}[\phi] 
 &= \pmb{1} + e^{\pm \beta \mu}\,\prod_\ell e^{\epsilon \kappa K} 
 e^{\pm \,i\sqrt{\epsilon U} \phi_\ell},
\end{align}
where $\phi_\ell$ is a diagonal matrix of the form 
$\phi_\ell=(\phi_{\ell,\bx}\,\delta_{\bx\by})$.
Note that, 
while the action is real-valued for the half-filling case $(\mu=0)$ 
due to the identity $M^b[\phi]|_{\mu=0}=(M^a[\phi]|_{\mu=0})^\ast$, 
it is generically complex-valued when $\mu\neq 0$.

The expectation values of such observables 
that are made solely from the number density operators 
$n_\bx \equiv n_{\bx,\uparrow}+n_{\bx,\downarrow}-1
=n^a_\bx-n^b_\bx$ 
can be evaluated as a path integral over $\phi$ 
by simply replacing $n_\bx$ by $(i\sqrt{\epsilon U})^{-1}\,\phi_{\ell=0,\bx}$,  
as easily proved by using the operator identity 
\begin{align}
 &\int d\phi\, e^{-(1/2)\,\phi^2 + \,i\sqrt{\epsilon U}\,\phi\,(n^a_\bx-n^b_\bx)}
 \,(n^a_\bx-n^b_\bx)
\nonumber
\\ 
 &= \int d\phi\, e^{-(1/2)\,\phi^2 
 + \,i\sqrt{\epsilon U}\,\phi\,(n^a_\bx-n^b_\bx)}
 \,\phi/(i\sqrt{\epsilon U}). 
\end{align}
For example, the expectation value of the number density operator, 
$\langle n\rangle_S$, 
can be rewritten to a path integral form as in \eqref{n_PI}. 


As for observables of general form, 
one can resort to the Wick-Bloch-de Dominicis theorem,
\begin{align}
 &{\rm tr}\,\bigl[ e^{\hat{A}}\,
 a_{\bx_m}\cdots a_{\bx_1}\,a^\dag_{\bx'_1} \cdots a^\dag_{\bx'_{m'}} \bigr]
\nonumber
\\
 &= \delta_{m m'}\,\det(1+e^A)\,
 \left|
 \begin{array}{ccc}
  (1+e^A)^{-1}_{\bx_1 \bx'_1} & \cdots 
  & (1+e^A)^{-1}_{\bx_1 \bx'_m} \\
  \vdots & \ddots & \vdots \\
  (1+e^A)^{-1}_{\bx_m \bx'_1} & \cdots 
  & (1+e^A)^{-1}_{\bx_m \bx'_m}
 \end{array} \right|,
\end{align}
to obtain the following expression:
\begin{widetext}
\begin{align}
 &\frac{{\rm tr}\,
 \bigl[(e^{-\epsilon H_1}\,e^{-\epsilon H_2})^{N_\tau}\,
 a_{\bx_m}\cdots a_{\bx_1}\,a^\dag_{\bx'_1} \cdots a^\dag_{\bx'_{m'}}\,
 b_{\by_n}\cdots b_{\by_1}\,b^\dag_{\by'_1} \cdots b^\dag_{\by'_{n'}}\bigr]}
 {{\rm tr}\,\bigl[ (e^{-\epsilon H_1}\,e^{-\epsilon H_2})^{N_\tau} \bigr]}
\nonumber
\\
 &= \frac{\delta_{m m'}\,\delta_{n n'}}{Z}\,
 \int[d\phi]\,e^{-S[\phi]}\,
 \left|
 \begin{array}{ccc}
  \Delta^a_{\bx_1 \bx'_1} & \cdots 
  & \Delta^a_{\bx_1 \bx'_m} \\
  \vdots & \ddots & \vdots \\
  \Delta^a_{\bx_m \bx'_1} & \cdots 
  & \Delta^a_{\bx_m \bx'_m}
 \end{array} \right| \cdot
 \left|
 \begin{array}{ccc}
  \Delta^b_{\by_1 \by'_1} & \cdots 
  & \Delta^b_{\by_1 \by'_n} \\
  \vdots & \ddots & \vdots \\
  \Delta^b_{\by_n \by'_1} & \cdots 
  & \Delta^b_{\by_n \by'_n}
 \end{array} \right|
\quad \Bigl( Z = \int[d\phi]\,e^{-S[\phi]} \Bigr), 
\label{BdD}
\end{align}
\end{widetext}
where $\Delta^{a/b}[\phi]=M^{a/b}[\phi]^{-1}$.

\section{Evaluation of the trace under the Trotter decomposition
\label{sec:Trotter}}
The Hilbert space $\mathbb{V}$ of the Hubbard model 
after the particle-hole transformation 
is the tensor product of two Fock spaces, 
$\mathbb{V} = \mathbb{V}^a \otimes \mathbb{V}^b$, 
each constructed by acting $a_\bx^\dag$ or $b_\bx^\dag$ 
on the Fock vacuum $|0\rangle$.
In this appendix, 
we give the explicit forms of the matrix elements 
that appear in the trace under the Trotter decomposition: 
\begin{align}
 \langle n \rangle_S
 = \frac{{\rm tr}\,\bigl[
 (e^{-\epsilon H_1}\,e^{-\epsilon H_2})^{N_\tau}\,n \bigr]}
 {{\rm tr}\,\bigl[
 (e^{-\epsilon H_1}\,e^{-\epsilon H_2})^{N_\tau} \bigr]}
 = \frac{{\rm tr}\,\bigl[
 (T_1 T_2)^{N_\tau}\,n \bigr]}
 {{\rm tr}\,\bigl[
 (T_1 T_2)^{N_\tau} \bigr]}.
\end{align}
Here, the one-body part $H_1$ and 
the two-body part $H_2$ of the Hamiltonian are given by 
[see \eqref{H1} and \eqref{H2}]
\begin{align}
 H_1 &= H^a_1 \otimes \pmb{1} + \pmb{1} \otimes H^b_1,
\\
 H^{a}_1 &= \sum_{\bx,\by}\,h^a_{\bx\by}\,a_\bx^\dag a_\by 
 \equiv {} 
 -\sum_{\bx,\by}\, (\kappa K + \mu\, \pmb{1})_{\bx\by}\,a_\bx^\dagger a_\by,
\\ 
 H^{b}_1 &= \sum_{\bx,\by}\,h^b_{\bx\by}\,b_\bx^\dag b_\by 
 \equiv {} 
 -\sum_{\bx,\by}\, (\kappa K - \mu\, \pmb{1})_{\bx\by}\,b_\bx^\dagger b_\by,
\\
 H_2 &={} -U\,\sum_\bx (n^a_\bx-1/2)\otimes(n^b_\bx-1/2). 
\end{align}
The number density operator is given by 
\begin{align}
 n &= \frac{1}{N_s}\,\sum_\bx 
 (n^a_\bx\otimes \pmb{1} - \pmb{1} \otimes n^b_\bx),
\end{align}
and we have introduced the transfer matrices corresponding to $H_1$ and $H_2$:
\begin{align}
 T_1 &\equiv e^{-\epsilon H_1} = e^{-\epsilon H^a_1}\otimes e^{-\epsilon H^b_1}
 \equiv T^a_1\otimes T^b_1,
\nonumber
\\
 T_2 &\equiv e^{-\epsilon H_2}.
\end{align}

We first introduce a one-dimensional ordering 
to the set of all spatial coordinates, $\Lambda=\{\bx\}$, 
and take a basis of $\mathbb{V}$ to be 
\begin{align}
 \{ | X \rangle \otimes | Y \rangle \},
\end{align}
where the states
\begin{align}
 |X\rangle &\equiv a_{\bx_1}^\dag a_{\bx_2}^\dag \cdots a_{\bx_m}^\dag |0\rangle
 \in \mathbb{V}^a,
\\
 |Y\rangle &\equiv b_{\by_1}^\dag b_{\by_2}^\dag \cdots b_{\by_n}^\dag |0\rangle
 \in \mathbb{V}^b, 
\end{align}
are labeled by the subsets of ordered coordinates, 
$X=\{\bx_1,\bx_2,\ldots,\bx_m\}\subset \Lambda$ 
(with $\bx_1<\bx_2<\cdots<\bx_m$), 
$Y=\{\by_1,\by_2,\ldots,\by_n\}\subset \Lambda$ 
(with $\by_1<\by_2<\cdots<\by_n$). 
We will denote their sizes by $|X|=m$, $|Y|=n$.

The matrix elements of $T^a_1 = e^{-\epsilon H^a_1}$ 
are then given by the following determinants:
\begin{align}
 &(T^a_1)_{X X'} = \left|
 \begin{array}{ccc}
  (e^{-\epsilon h^a})_{\bx_1 \bx_1'} & \cdots 
  & (e^{-\epsilon h^a})_{\bx_1 \bx_m'} \\
  \vdots & \ddots & \vdots \\
  (e^{-\epsilon h^a})_{\bx_m \bx_1'} & \cdots 
  & (e^{-\epsilon h^a})_{\bx_m \bx_m'}
 \end{array} \right| \,\delta_{|X|,|X'|}
\nonumber
\\
 &= e^{\epsilon\mu |X|}\,\left|
 \begin{array}{ccc}
  (e^{\epsilon \kappa K})_{\bx_1 \bx_1'} & \cdots 
  & (e^{\epsilon \kappa K})_{\bx_1 \bx_m'} \\
  \vdots & \ddots & \vdots \\
  (e^{\epsilon \kappa K})_{\bx_m \bx_1'} & \cdots 
  & (e^{\epsilon \kappa K})_{\bx_m \bx_m'}
 \end{array} \right| \,\delta_{|X|,|X'|}
\nonumber
\\
 & \qquad(m \equiv |X|=|X'|),
\end{align}
as can be easily proven 
by investigating the action of $T^a_1$ on the state $|X'\rangle$:
\begin{align}
 T^a_1\,|X'\rangle 
 &= e^{-\epsilon \sum_{\bx\by} h^a_{\bx\by}  a_\bx^\dag a_\by}\,
 a_{\bx_1'}^\dag\cdots a_{\bx_m'}^\dag |0\rangle
\nonumber
\\
 &\equiv \sum_X |X\rangle \, (T^a_1)_{X X'},
\end{align}
where the coefficients do not vanish 
only when $|X|=|X'|\,(=m)$. 
The matrix elements of $T^b_1 = e^{-\epsilon H^b_1}$ 
can also be given in the forms of determinant,
\begin{align}
 &(T^b_1)_{Y Y'} 
\nonumber
\\
 &= e^{-\epsilon\mu |Y|}\,\left|
 \begin{array}{ccc}
  (e^{\epsilon \kappa K})_{\by_1 \by_1'} & \cdots 
  & (e^{\epsilon \kappa K})_{\by_1 \by_n'} \\
  \vdots & \ddots & \vdots \\
  (e^{\epsilon \kappa K})_{\by_n \by_1'} & \cdots 
  & (e^{\epsilon \kappa K})_{\by_n \by_n'}
 \end{array} \right| \,\delta_{|Y|,|Y'|}
\nonumber
\\
 & \qquad(n \equiv |Y|=|Y'|).
\end{align}
We thus obtain the explicit forms of the matrix elements 
$(T_1)_{XY,X'Y'} = (T^a_1)_{X X'}\,(T^b_1)_{Y Y'}$.

As for $T_2 = e^{-\epsilon H_2}$, 
we note that $H_2$ acts on $|X\rangle \otimes |Y\rangle$ diagonally: 
\begin{align}
 &H_2\,|X\rangle \otimes |Y\rangle
\nonumber
\\
 &={}-U\,\sum_\bz\, \Bigl(a_\bz^\dag a_\bz-\frac{1}{2}\Bigr) |X\rangle \otimes
 \Bigl(b_\bz^\dag b_\bz-\frac{1}{2}\Bigr) |Y\rangle
\nonumber
\\
 &\equiv (h_2)_{X Y}\,|X\rangle \otimes |Y\rangle.
\end{align}
The coefficients $(h_2)_{X Y}$ can be calculated easily to be
\begin{align}
 &(h_2)_{X Y}
\nonumber
\\
 &= {}-\frac{U}{4}\,\sum_\bz\,
 \bigl[ \theta(\bz\in X)\,\theta(\bz\in Y) 
 - \theta(\bz\in X)\,\theta(\bz\notin Y)
\nonumber
\\
 &~~~~{} - \theta(\bz\notin X)\,\theta(\bz\in Y)
 + \theta(\bz\notin X)\,\theta(\bz\notin Y) \bigr]
\nonumber
\\
 &={} -\frac{U}{4}\,( 2|X\cap Y| + 2|\bar{X}\cap \bar{Y}| - N_s ), 
\end{align}
where $\theta()$ is the logical step function 
and $\bar{X}$ stands for the complement of the set $X$, 
$\bar{X}=\Lambda\setminus X$. 
The matrix elements of $T_2$ is then given by 
$(T_2)_{XY,X'Y'} = e^{-\epsilon (h_2)_{X Y}}\,\delta_{X X'}\,\delta_{Y Y'}$.

Finally, the matrix elements of $n$ are given by
\begin{align}
 n_{XY,X'Y'} = \frac{1}{N_s}\,(|X|-|Y|)\,\delta_{X X'}\,\delta_{Y Y'}. 
\end{align}
With the matrix elements given above, 
$\langle n \rangle_S$ can be expressed as 
\begin{align}
 \langle n \rangle_S = \frac{1}{N_s}\,
 \frac{\sum_{X,Y\subset\Lambda} \bigl[(T_1\,T_2)^{N_\tau}\bigr]_{XY,XY}\,
 (|X|-|Y|)}
 {\sum_{X,Y\subset\Lambda} \bigl[(T_1\,T_2)^{N_\tau}\bigr]_{XY,XY}}.
\end{align}

\section{Summary of the parameters in the computation
\label{sec:parameters}}
We summarize the parameters relevant to the TLTM 
in the estimation of $\langle n \rangle_S$. 
We order the termination times $t_a$ for replicas $a$ 
as $t_0=0 < t_1 < \cdots < t_A=T$ ($T$: the largest flow time), 
and set $t_a$ to be a piecewise linear function of $a$ 
with a single breakpoint at $a=a_c$, 
by assuming that 
the deformed region reaches the vicinity 
of all the relevant Lefschetz thimbles at almost the same flow time 
and that the linear form is effective also for the transient period: 
\begin{align}
 t_a = 
 \left\{
 \begin{array}{cl}
  t_c\,a/a_c & (0 \leq a \leq a_c)\\
  t_c + (T-t_c)\,(a-a_c)/(A-a_c) & (a_c < a \leq A)
 \end{array}
 \right..
\label{piecewise_linear}
\end{align}
Each Monte Carlo step consists of 50 Metropolis tests in the $x$ direction 
and $N_{\rm swap}$ swaps of configurations at adjacent replicas, 
and the flow equations \eqref{eq:flow1} and \eqref{eq:flow2} 
are integrated numerically by using the adaptive Runge-Kutta of 7-8th order. 
For each value of $\beta\mu$, 
we make a test run with small statistics 
and adjust the parameters $A$, $t_c$, $a_c$ 
in such a way that the acceptance rates of the swapping process 
at adjacent replicas are almost the same for all pairs 
(being roughly above 40\%). 
After this, we make another test run of 1,000 data points 
to adjust the width of the Gaussian proposal 
in the Metropolis test in the $x$ direction 
so that the acceptance rate is in the range 50\%--80\%. 
This width varies depending on replicas $a$ and the values of $\beta\mu$. 
Using the adjusted parameters, 
we get a sample of size $N_{\rm conf}$ 
after discarding 5,000 configurations, 
and analyze the data by using the Jackknife method, 
with bins 
whose sizes are adjusted by taking account of autocorrelations. 
Finally, from the obtained data $\{\bar{n}_a\}$ 
$(a=a_{\rm min},\ldots,a_{\rm max}(=A))$ 
[see \eqref{eq:estimate}], 
we estimate the expectation value $\langle n \rangle_S$ 
by using the $\chi^2$ fit with a constant function of $a$. 
We confirm that the system is in global equilibrium 
and the sample size is sufficient 
by looking at the optimized value of $\chi^2/{\rm DOF}$ 
with ${\rm DOF}=a_{\rm max}-a_{\rm min}$. 
The obtained results are summarized in Table \ref{table:TLTM_parameters}. 
\begin{table}[ht]
\centering
\begin{tabular}{|c||c|c|c|c|c|c|c|c|}
 \hline
 $\beta\mu$ & 1 & 2 & 3 & 4 & 5 & 6 & 7 & 8 \\
 \hline\hline
 $T/(\beta\mu)$ & 1/10 & 1/10 & 1/10 & 1/10 & 1/10 & 1/11 & 1/11 & 1/12 \\
 \hline
 $A$  & 8 & 9 & 10 & 11 & 11 & 11 & 11 & 11 \\
 \hline
 $t_c/T$ & 0.7 & 0.6 & 0.6 & 0.6 & 0.6 & 0.5 & 0.5 & 0.5  \\
 \hline
 $a_c$ & 5 & 5 & 6 & 6 & 6 & 5 & 5 & 5  \\
 \hline
 $N_{\rm swap}$  & 10 & 10 & 12 & 12 & 12 & 12 & 12 & 12 \\
 \hline
 $N_{\rm conf}$ & 5k & 5k & 10k & 10k & 15k & 25k & 15k & 15k  \\
 \hline
 $a_{\rm min}$  & 0 & 0 & 2 & 3 & 5 & 5 & 6 & 7 \\
 \hline
 $\chi^2/{\rm DOF}$ & 0.53 & 0.43 & 0.47 & 0.12 & 0.45 & 0.39 & 1.07 & 0.72  \\
 \hline
\end{tabular}\\
\vspace{5mm}
\begin{tabular}{|c||c|c|c|c|c|c|c|c|}
 \hline
 $\beta\mu$ & 9 & 10 & 11 & 12 & 13 & 14 & 15 & 16 \\
 \hline\hline
 $T/(\beta\mu)$   & 1/12 & 1/12 & 1/12 & 1/11 & 1/11 & 1/11 & 1/11 & 1/11  \\
 \hline
 $A$   & 11 & 11 & 12 & 12 & 12 & 12 & 12 & 12  \\
 \hline
 $t_c/T$ & 0.5 & 0.55 & 0.55 & 0.6 & 0.6 & 0.6 & 0.6 & 0.6   \\
 \hline
 $a_c$ & 5 & 5 & 6 & 6 & 6 & 7 & 7 & 7   \\
 \hline
 $N_{\rm swap}$ & 12 & 12 & 14 & 14 & 14 & 14 & 14 & 14  \\
 \hline
 $N_{\rm conf}$ & 10k & 10k & 10k & 10k & 5k & 5k & 5k & 5k   \\
 \hline
 $a_{\rm min}$ & 8 & 8 & 10 & 8 & 9 & 10 & 9 & 9  \\
 \hline
 $\chi^2/{\rm DOF}$ & 0.09 & 0.92 & 0.21 & 1.74 & 0.40 & 0.17 & 0.75 & 0.20  \\
 \hline
\end{tabular}
\caption{TLTM parameters and the results}
\label{table:TLTM_parameters}
\end{table}
%

\section{More on the fine-tuning of flow time without tempering
\label{sec:fine-tuning}}
In order to understand the difficulty 
to find such an intermediate value of flow time 
that avoids both the sign and multimodal problems 
(without tempering), 
let us see the right panel of Fig.~\ref{fig:mu5_gltm}, 
which is the counterpart of Fig.~\ref{fig:mu5_tltm} (with tempering) 
for the same $\beta\mu=5$. 
We see that the estimated values have large statistical errors 
at smaller flow times (due to the sign problem) 
while they have small statistical errors around incorrect values
at larger flow times 
(due to the trapping of configurations at a small number of thimbles). 
The best flow time must be at the boundary between the two regions, 
but it should be a difficult task to find such value 
out of the set of flow times with finite spacings. 
In fact, if one takes a flow time from the smaller region, 
then, although the obtained estimate may happen to be close 
to the correct value, 
it must have a large statistical error. 
On the other hand, if a flow time is taken from the larger region, 
it will give an incorrect value (but with a small statistical error 
because only a small number of thimbles are sampled).

\begin{figure}[ht]
\includegraphics[width=28mm]{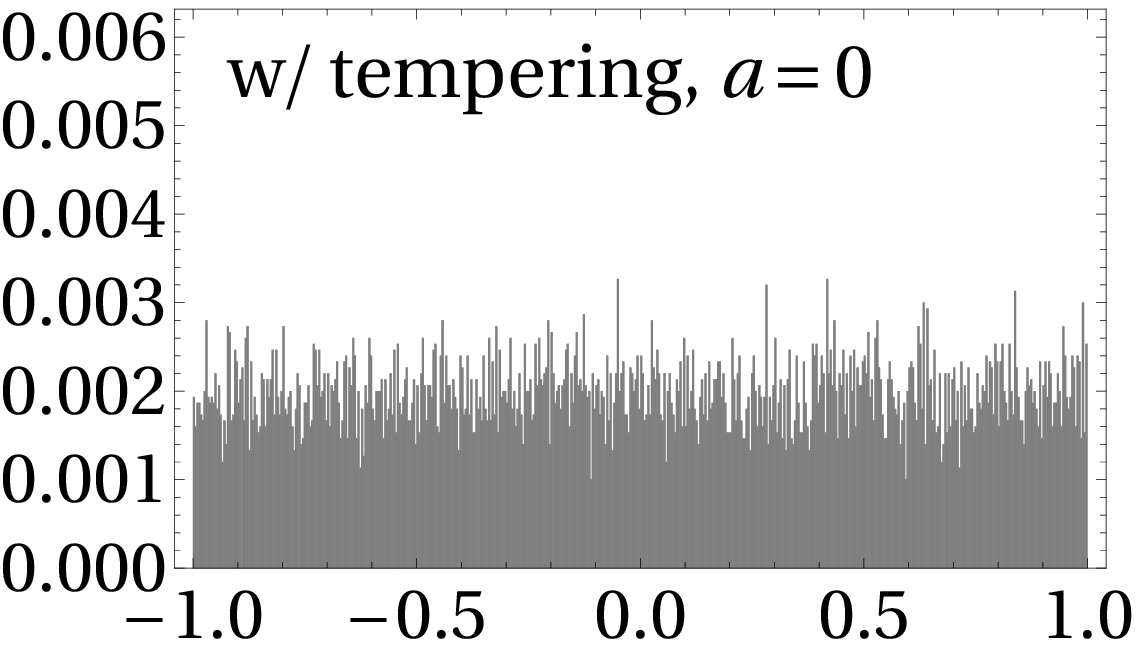} 
\includegraphics[width=28mm]{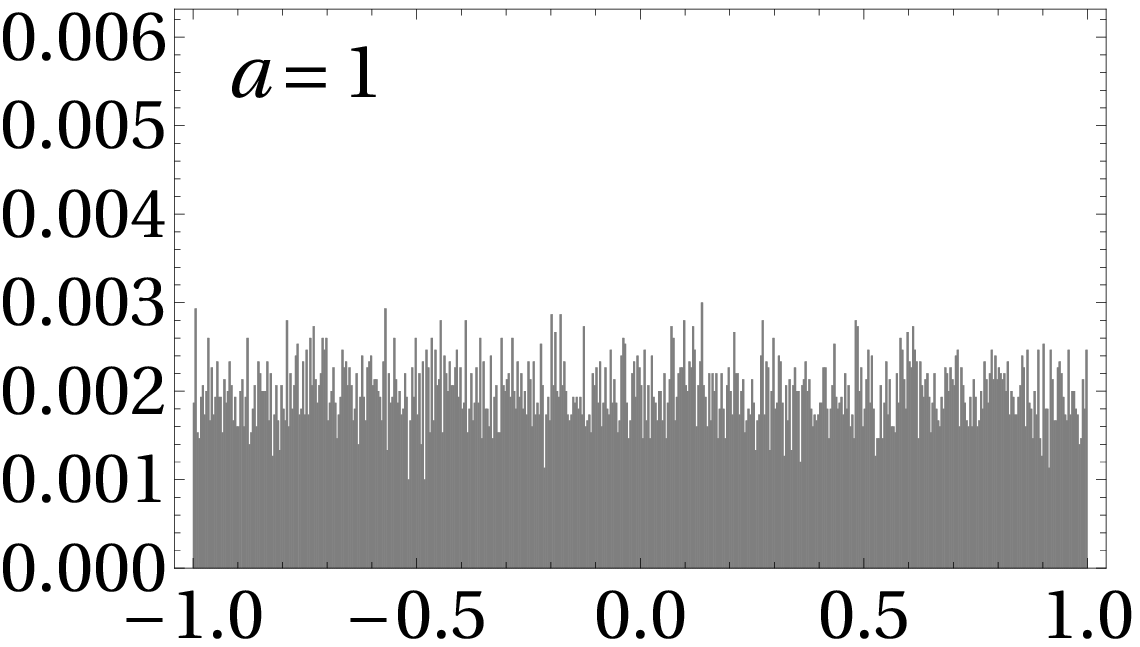} 
\includegraphics[width=28mm]{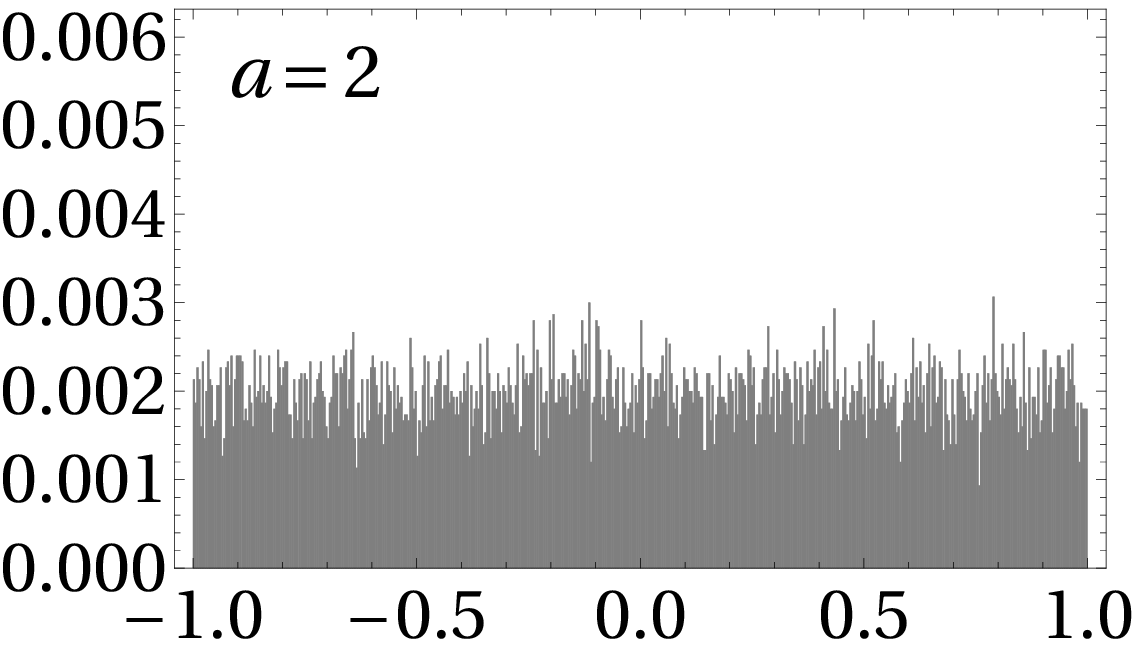} \\
\includegraphics[width=28mm]{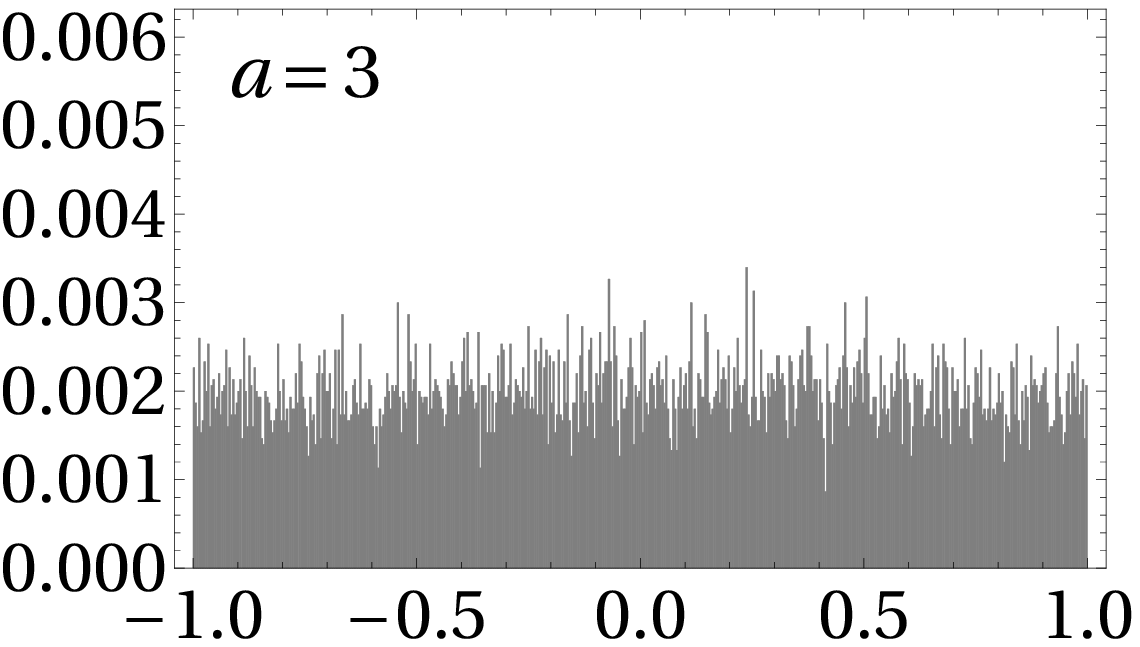} 
\includegraphics[width=28mm]{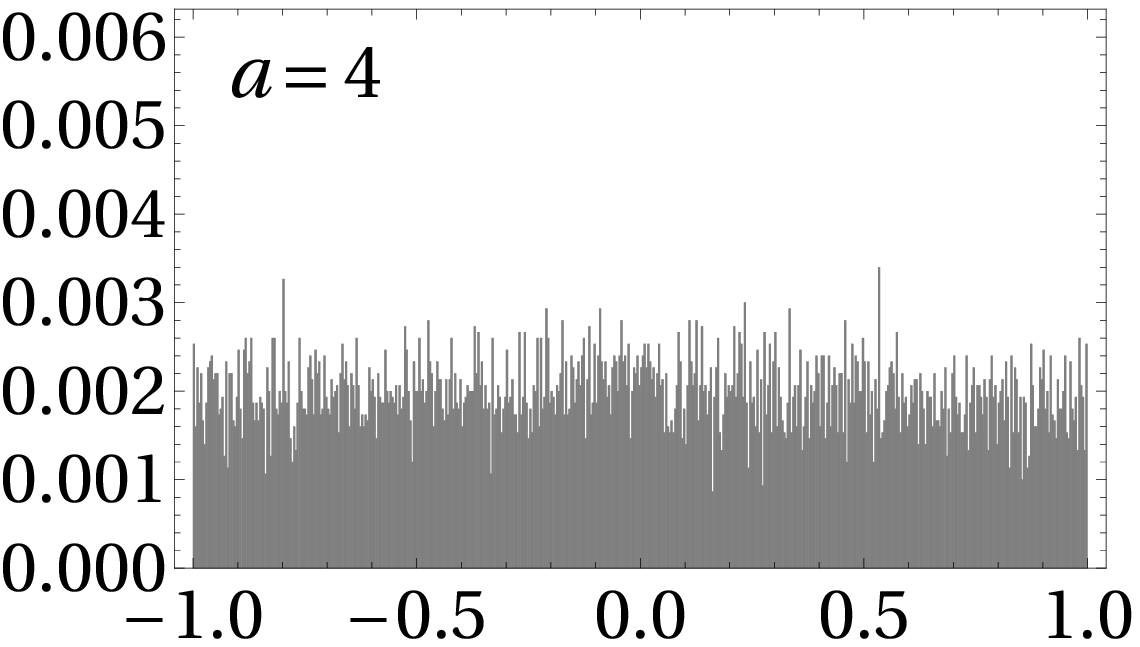} 
\includegraphics[width=28mm]{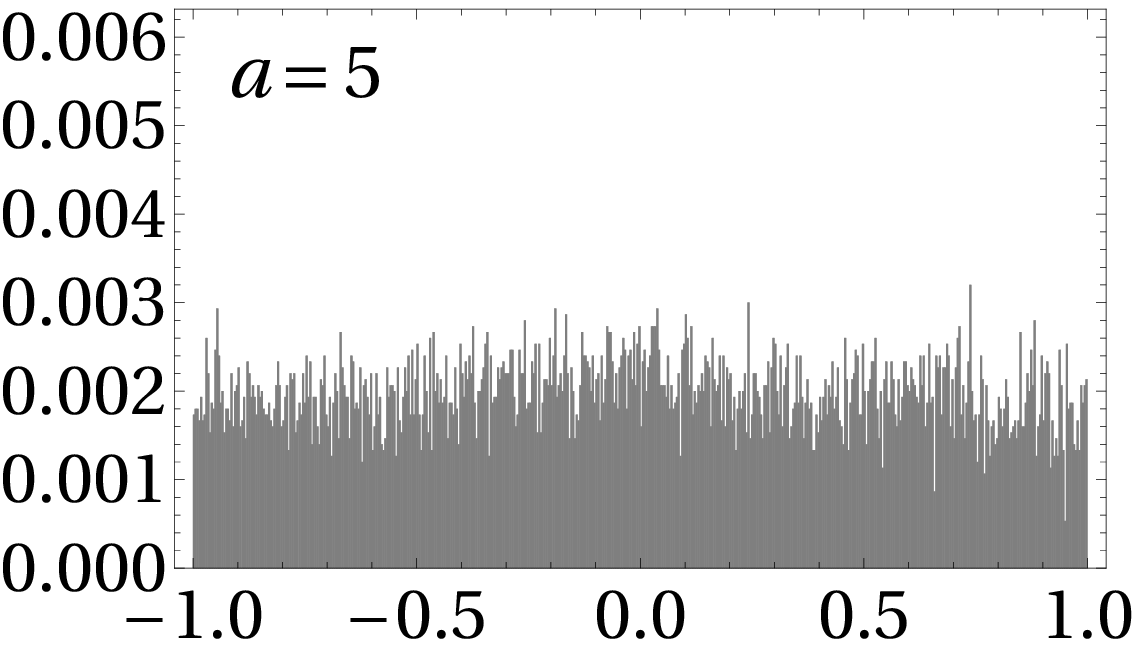} \\
\includegraphics[width=28mm]{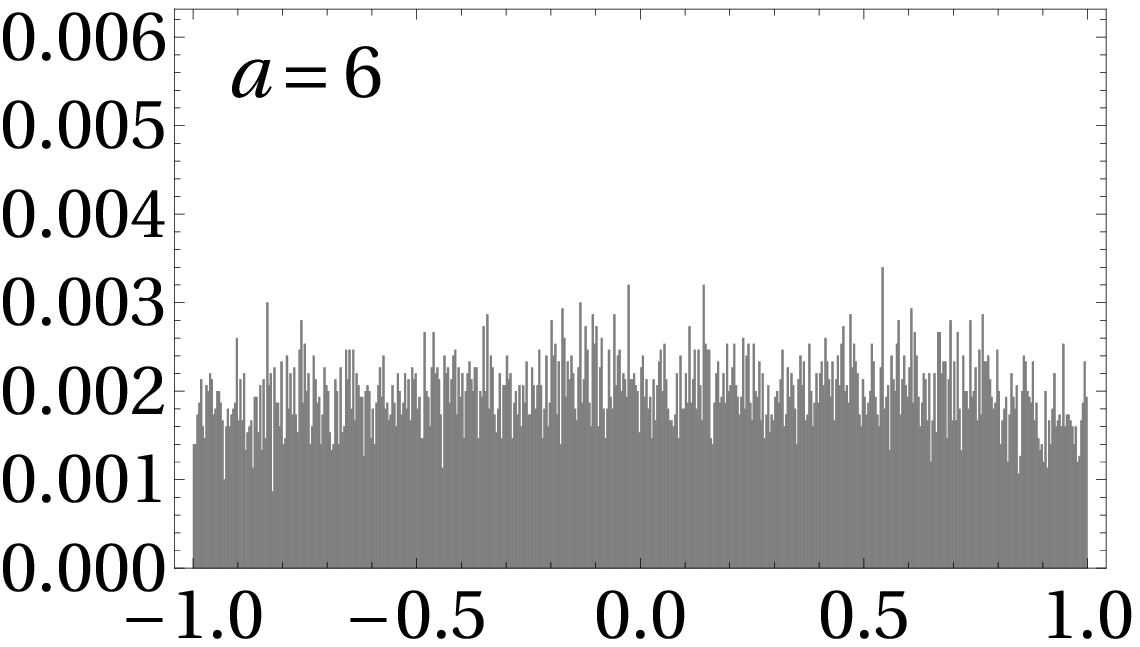} 
\includegraphics[width=28mm]{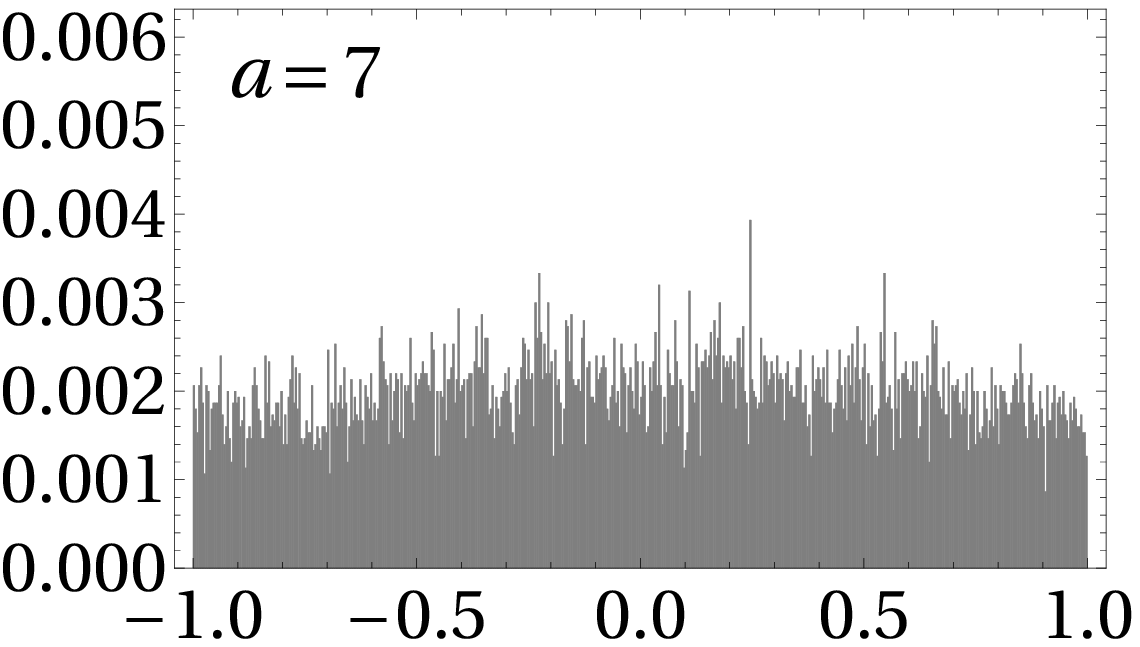} 
\includegraphics[width=28mm]{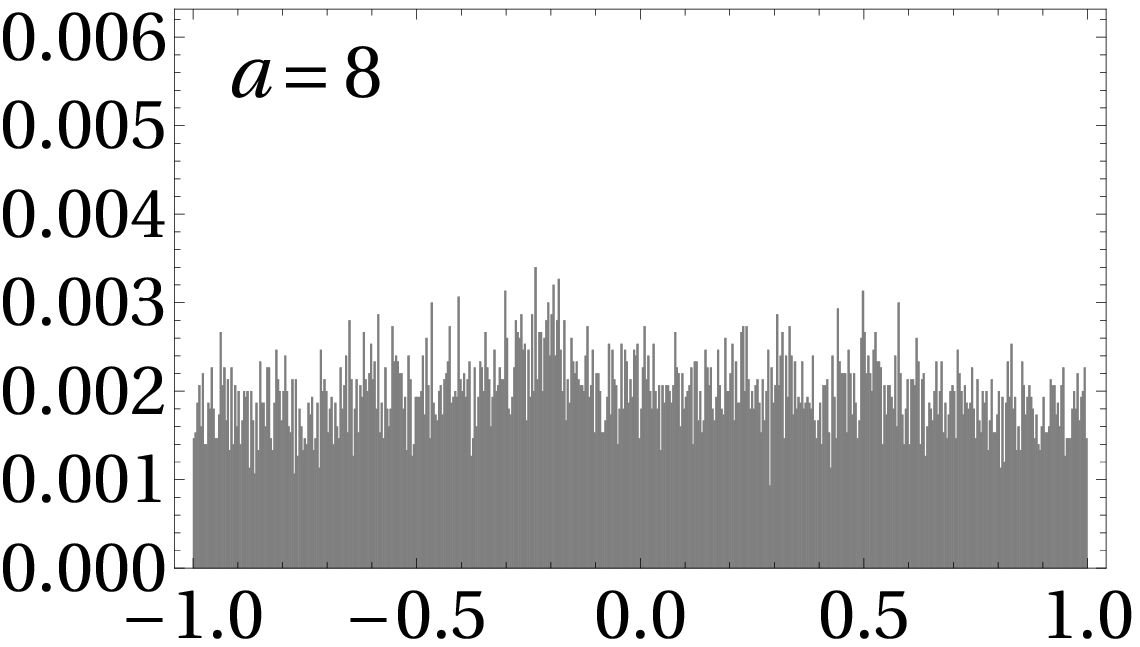} \\ 
\includegraphics[width=28mm]{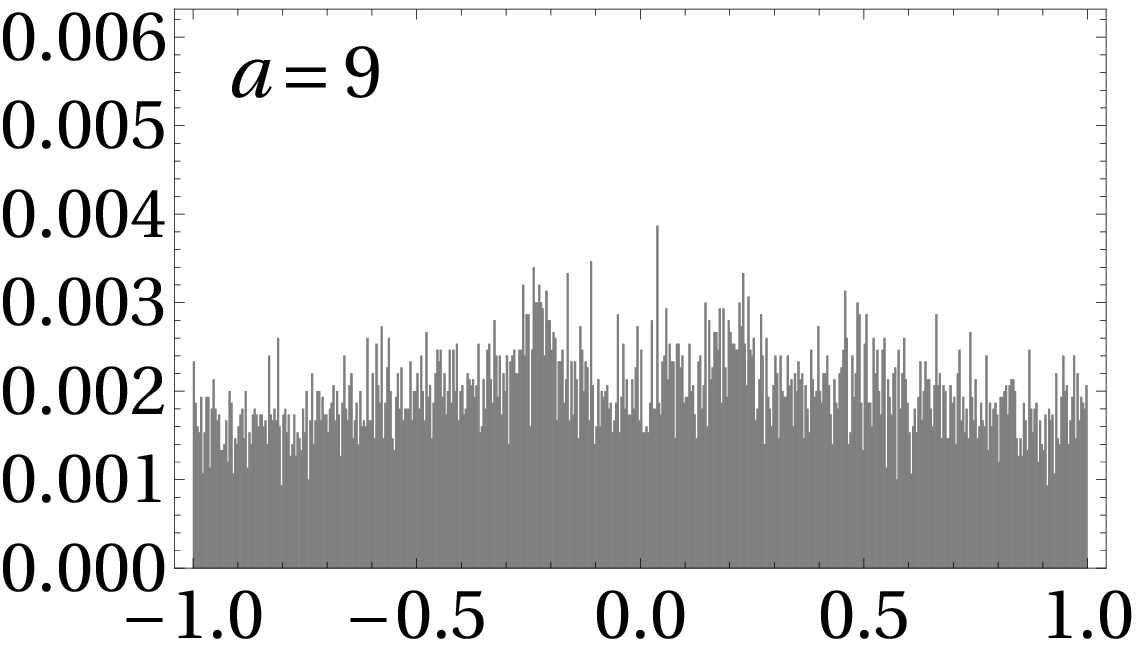} 
\includegraphics[width=28mm]{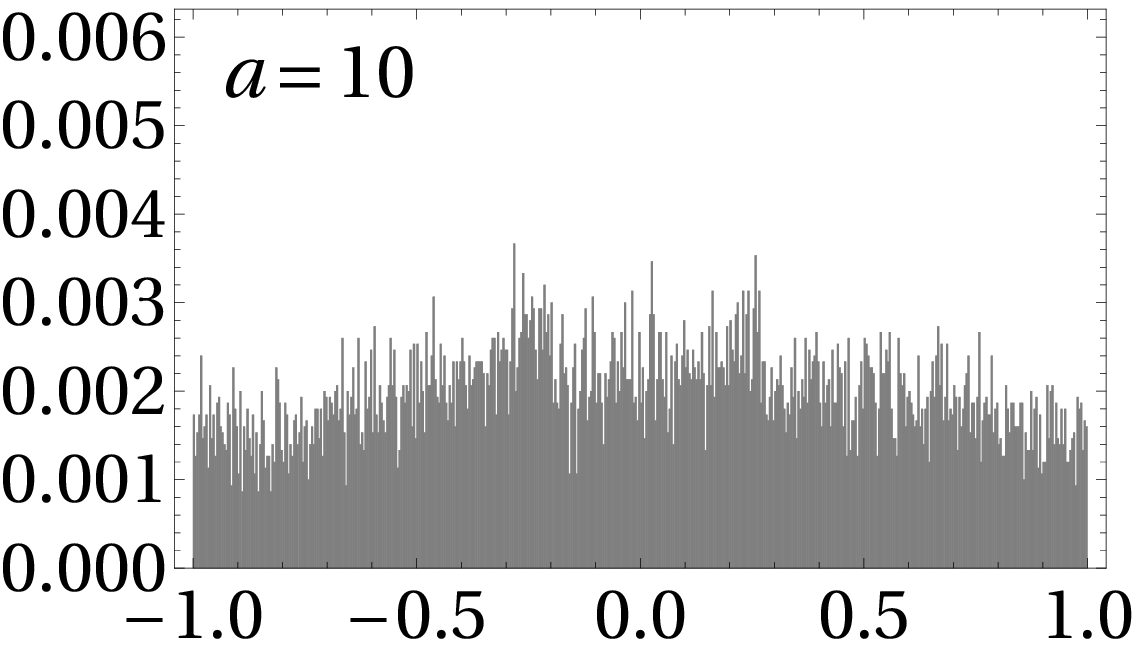} 
\includegraphics[width=28mm]{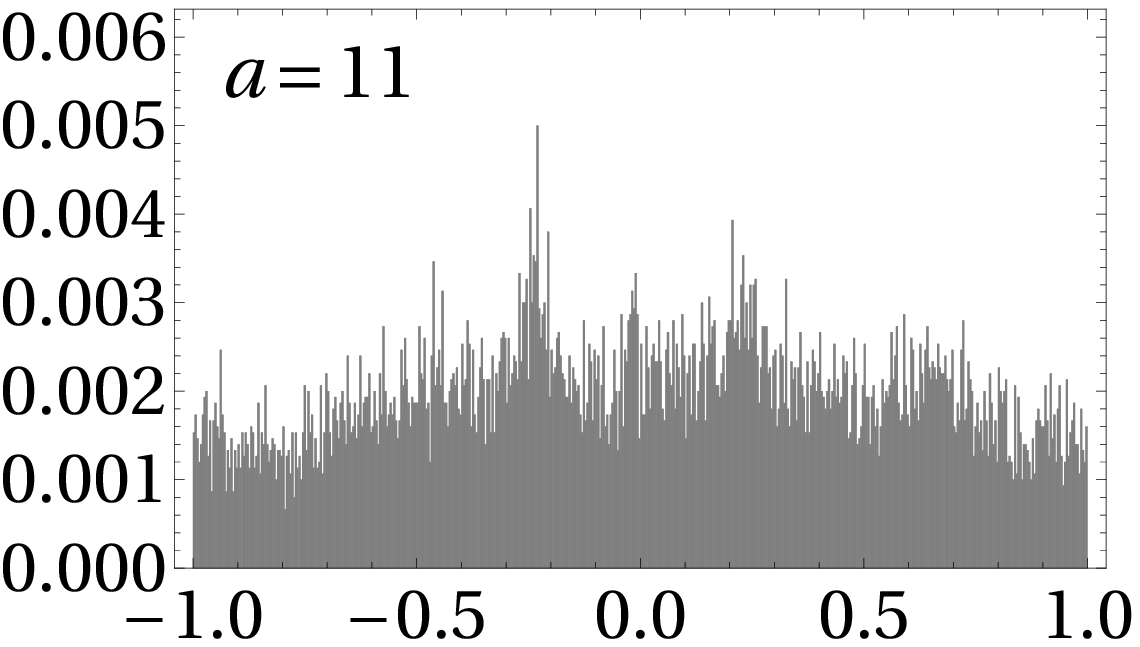} \\ 
\vspace{7mm}
\includegraphics[width=28mm]{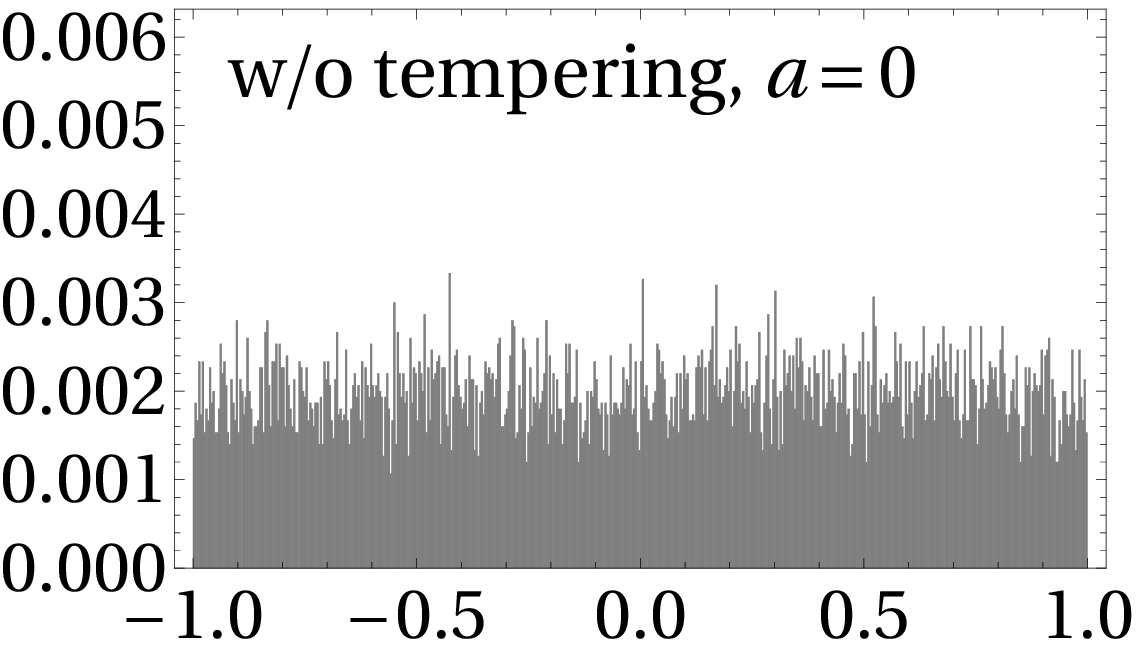} 
\includegraphics[width=28mm]{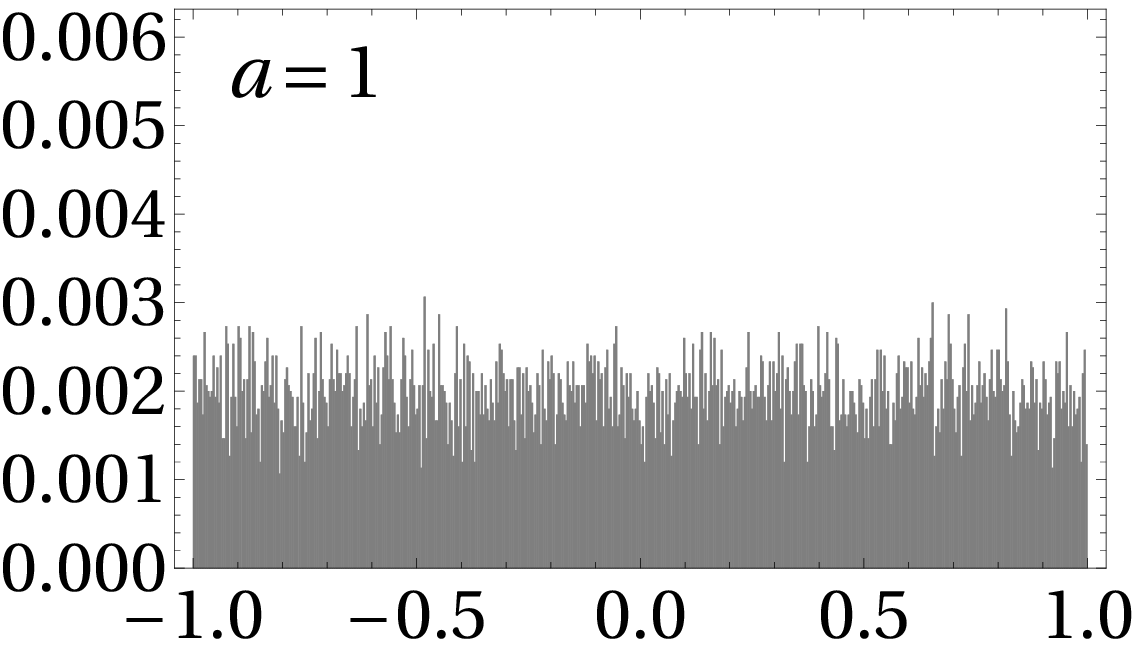} 
\includegraphics[width=28mm]{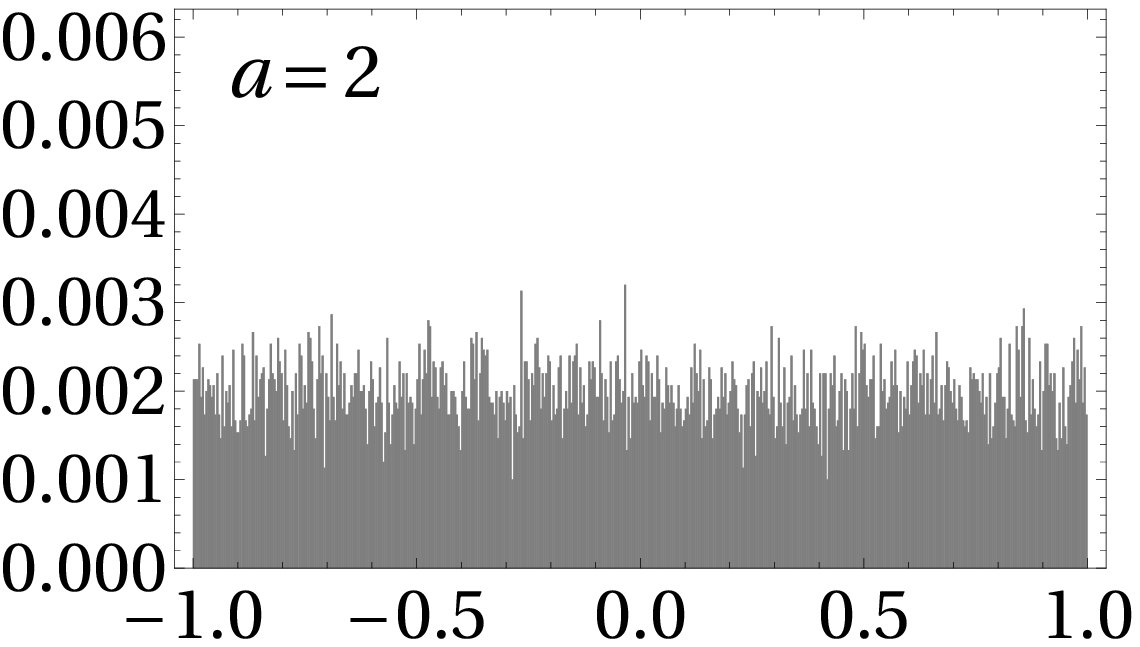} \\
\includegraphics[width=28mm]{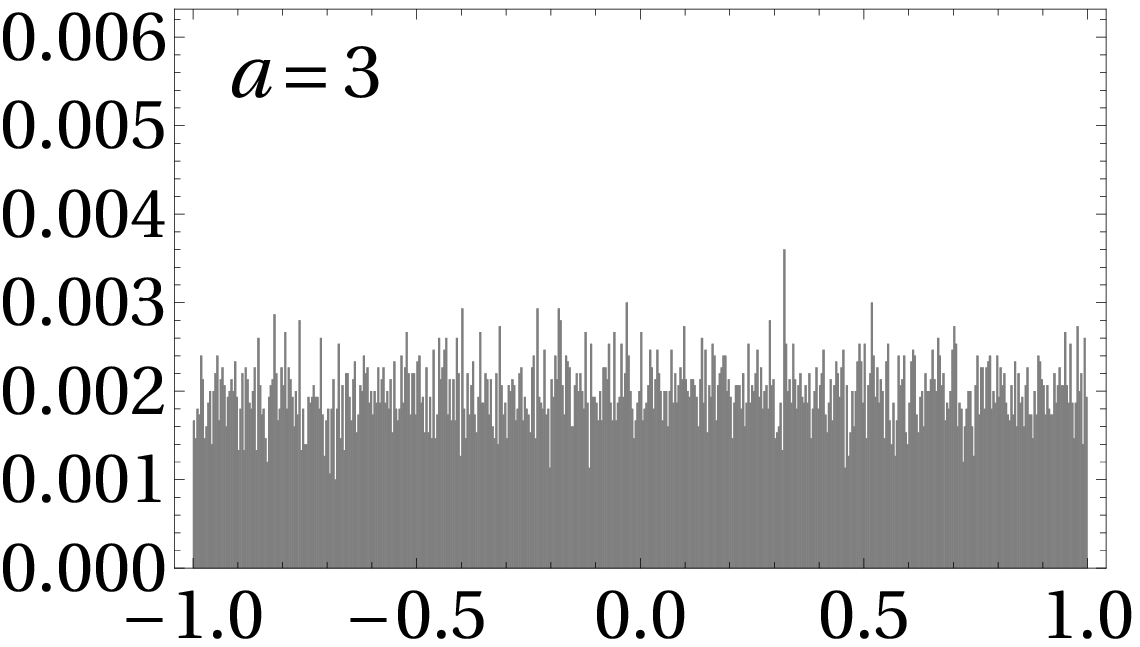} 
\includegraphics[width=28mm]{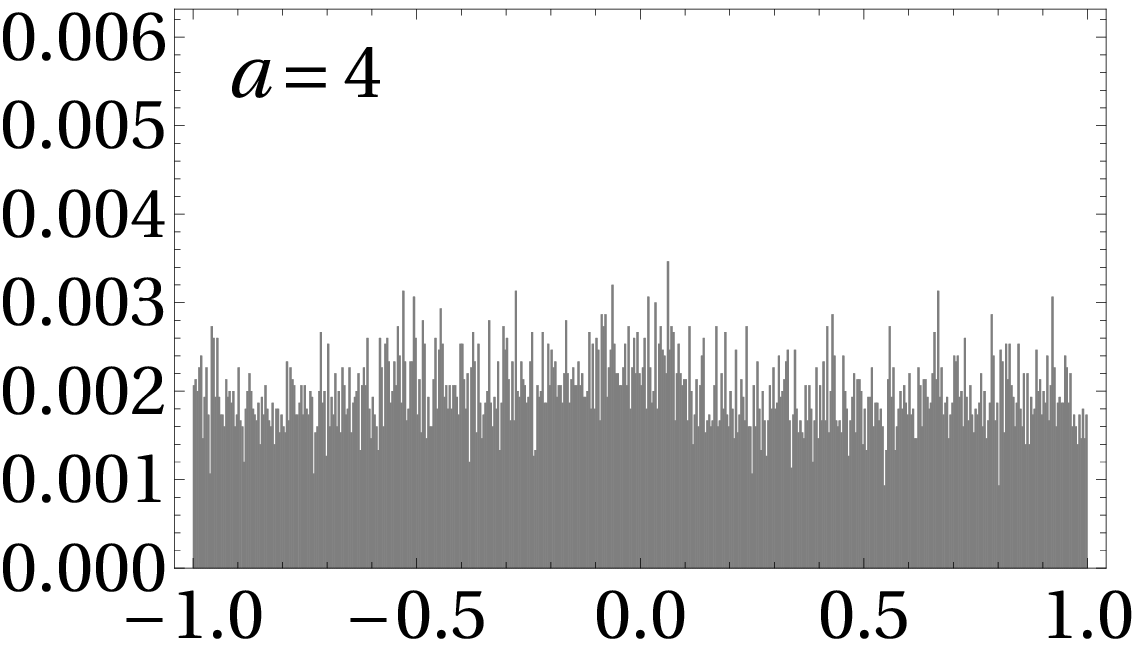} 
\includegraphics[width=28mm]{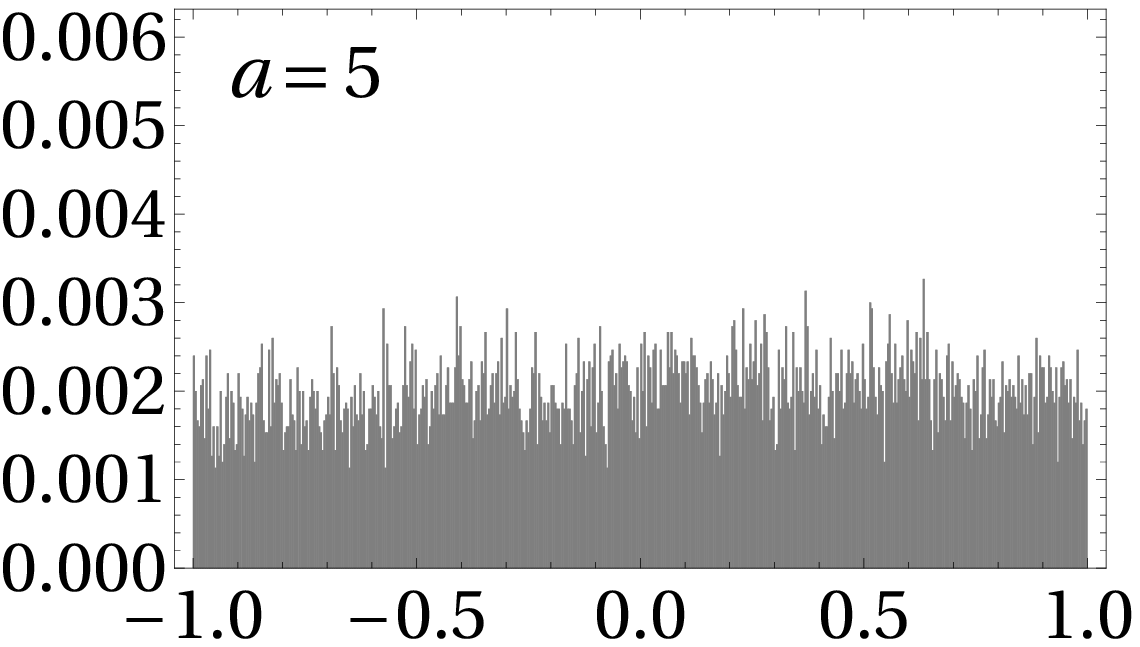} \\
\includegraphics[width=28mm]{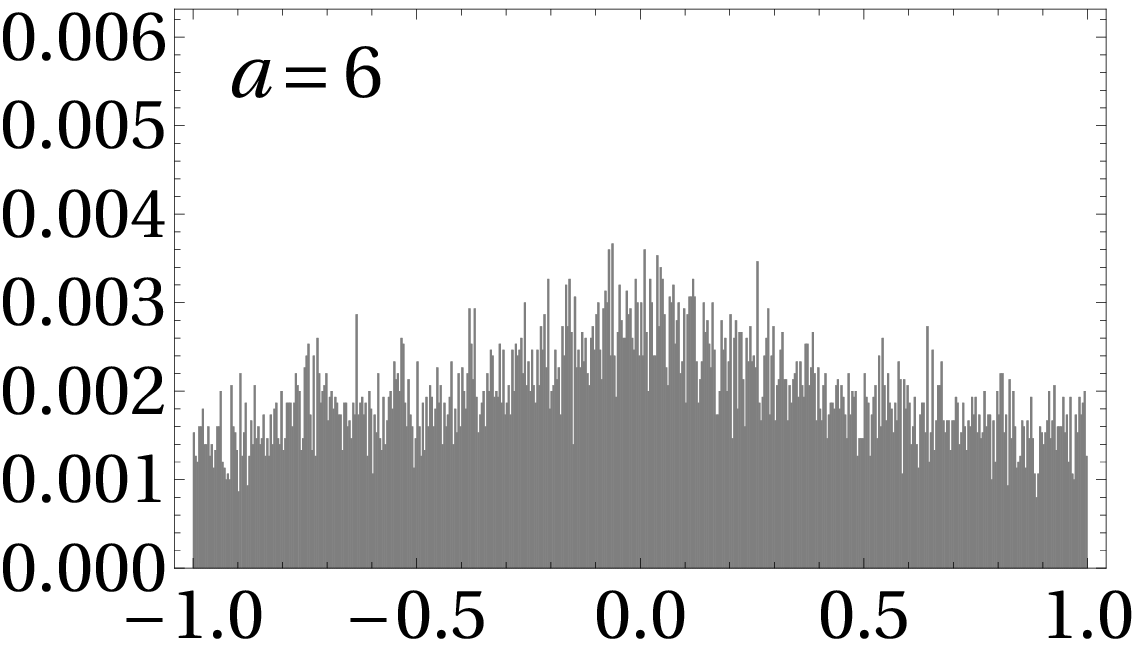} 
\includegraphics[width=28mm]{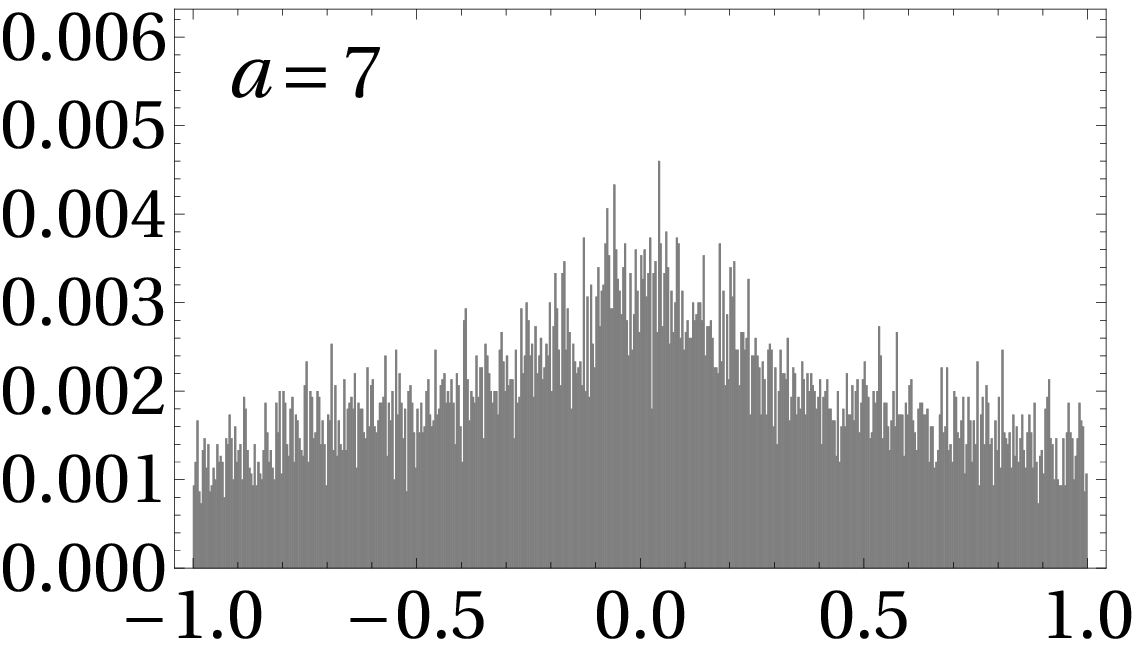} 
\includegraphics[width=28mm]{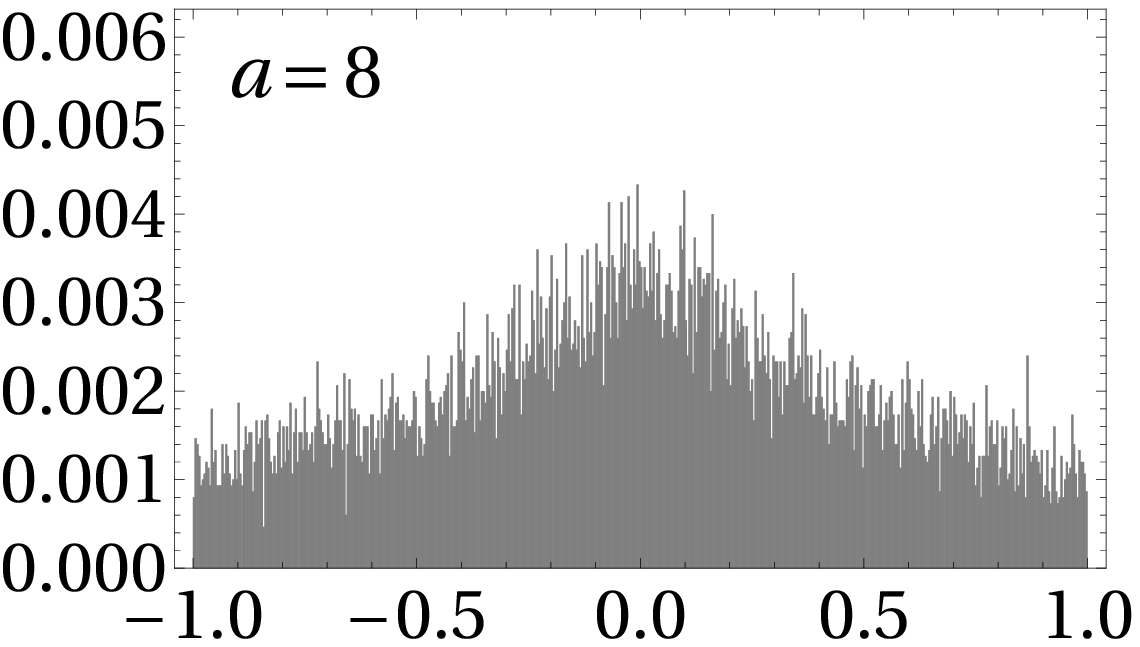} \\ 
\includegraphics[width=28mm]{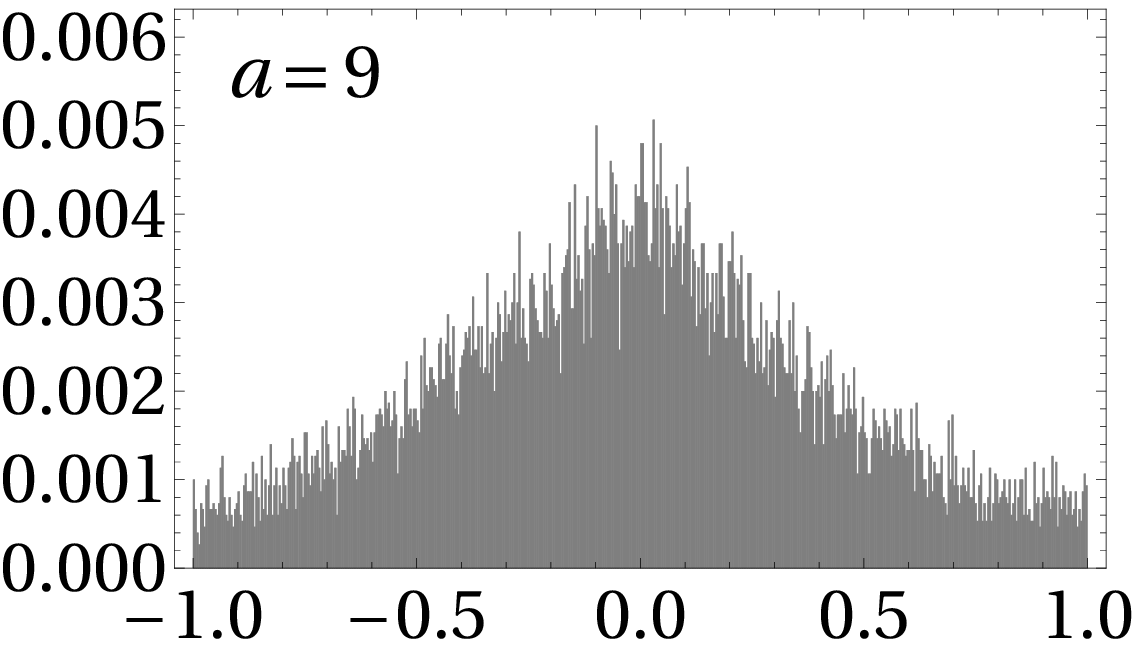} 
\includegraphics[width=28mm]{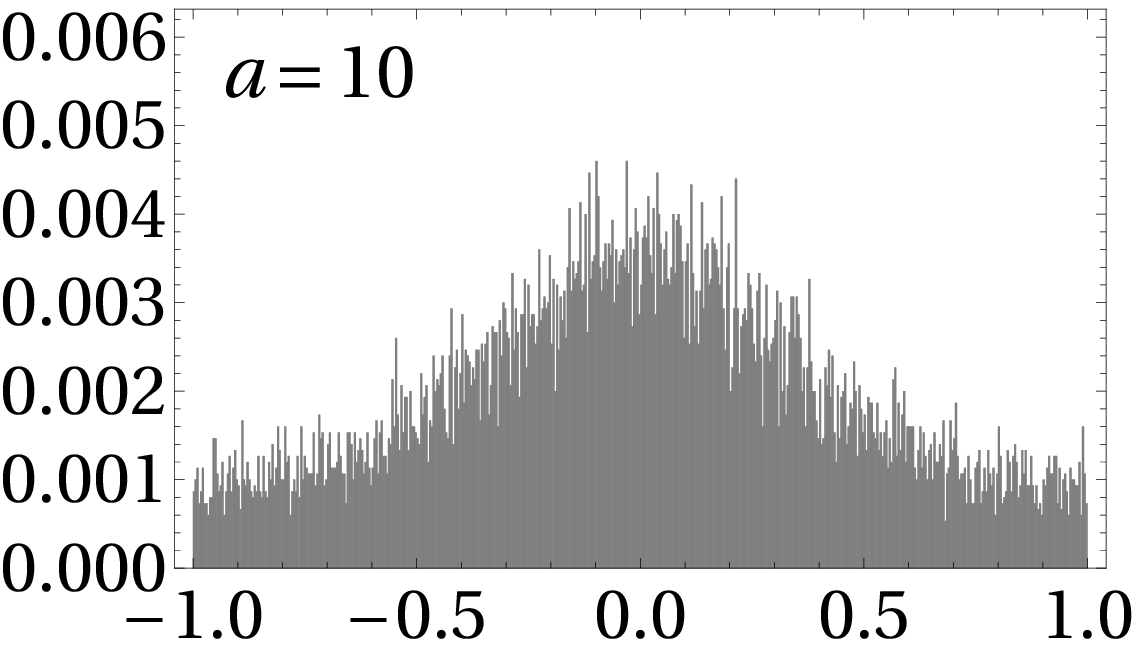} 
\includegraphics[width=28mm]{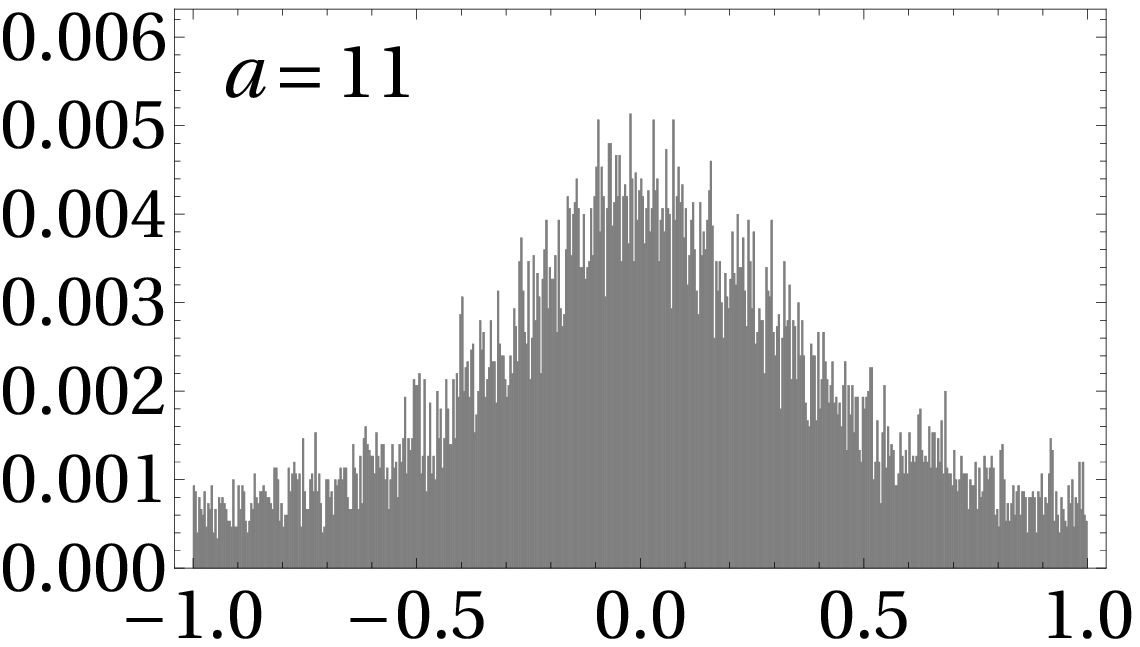} \\
\caption{
\label{fig:histo}
 Normalized histograms of $\theta_{t_a}(x)/\pi$ for $\beta\mu=5$. 
 (Top) with tempering. (Bottom) without tempering.
}
\end{figure}

In order to understand Fig.~\ref{fig:mu5_tltm} and Fig.~\ref{fig:mu5_gltm} 
as reflecting the extent of the sign and multimodal problems, 
let us see Fig.~\ref{fig:histo}, 
which depicts the normalized histograms of phases $\theta_{t_a}(x)$ 
for $\beta\mu=5$ 
with tempering (top) and without tempering (bottom). 
We see that at smaller flow times 
the histograms are almost flat for the both cases 
(giving rise to the sign problem), 
but at larger flow times 
those without tempering become almost unimodal 
(reflecting the trapping at a small number of thimbles) 
while those with tempering correctly come to have various peaks 
(which may not be so obvious from the figure 
because there are many peaks 
and each peak is broadened by the Jacobian determinant).



\end{document}